\documentclass[reprint,notitlepage,twocolumn,
superscriptaddress,aps,longbibliography]{revtex4-1}
\usepackage[utf8]{inputenc}
\usepackage{braket}
\usepackage{amsmath}
\usepackage{amsthm}
\usepackage{mathrsfs}
\usepackage{mathtools}
\usepackage{amssymb}
\usepackage{dsfont}
\usepackage{braket}
\usepackage{bm}
\DeclareMathOperator{\tr}{tr}
\usepackage{color}
\pdfoutput=1

\usepackage[english]{babel}
\usepackage[T1]{fontenc}
\usepackage{amsmath}
\usepackage[colorlinks=true]{hyperref}
\usepackage{enumitem}
\usepackage{amsthm}
\usepackage{amssymb}
\usepackage{graphicx}
\usepackage{float}
\usepackage[ruled]{algorithm2e}

\begin{document}

\title{
Quantum Process Learning Through Neural Emulation}
\author{Yan Zhu}
\affiliation{QICI Quantum Information and Computation Initiative, Department of Computer Science,
The University of Hong Kong, Pokfulam Road, Hong Kong}
\author{Ya-Dong Wu}
\email{yadongwu@hku.hk}
\thanks{Y Zhu and Y-D Wu contribute equally}
\affiliation{QICI Quantum Information and Computation Initiative, Department of Computer Science,
The University of Hong Kong, Pokfulam Road, Hong Kong}
\author{Qiushi Liu}
\affiliation{QICI Quantum Information and Computation Initiative, Department of Computer Science,
The University of Hong Kong, Pokfulam Road, Hong Kong}
\author{Yuexuan Wang}
\affiliation{AI Technology Lab, Department of Computer Science,
The University of Hong Kong, Pokfulam Road, Hong Kong}
\affiliation{ College of Computer Science and Technology,
Zhejiang University, Zhejiang Province, China}
\author{Giulio Chiribella }
\email{giulio@cs.hku.hk}
\affiliation{QICI Quantum Information and Computation Initiative, Department of Computer Science,
The University of Hong Kong, Pokfulam Road, Hong Kong}
\affiliation{Department of Computer Science, Parks Road, Oxford, OX1 3QD, United Kingdom}
\affiliation{Perimeter Institute for Theoretical Physics, Waterloo, Ontario N2L 2Y5, Canada}

\begin{abstract}
    Neural networks are a promising tool for    characterizing  intermediate-scale quantum devices from limited amounts of measurement data.
    A challenging  problem in this area is to learn the action of an unknown quantum process on an ensemble of physically relevant input states.  
     To tackle this problem, we introduce a neural network that emulates the unknown process by constructing an internal representation  of the input ensemble and by  mimicking the action of the process at the state representation level. 
    After being trained with  measurement data from  a few  pairs of input/output quantum states, the network becomes able to  predict the  measurement statistics for all inputs in the ensemble of interest.
We show that  our model exhibits high accuracy in  applications to quantum computing, quantum photonics, and quantum many-body physics. 
\end{abstract}

\maketitle

{\em Introduction.}  
Characterizing intermediate-scale quantum processes is a major challenge in quantum computing, due to the exponential blow-up of the number of samples required by conventional tomographic methods~\cite{chuang1997,dariano2001,altepeter2003,lobino2008,rahimi2011}.  
 To address this problem, alternative methods have been  proposed over the years.  Quantum shadow tomography~\cite{
 aaronson2018,huang2020}, recently extended from states to processes \cite{huang2022},  can  accurately characterize the action of 
 an unknown quantum process on input states that are invariant under local Clifford gates. 
  On the other hand, neural networks have been used to predict the evolution of a single input state subject to an unknown quantum dynamics~\cite{flurin2020,mohseni2022,koolstra2022,huangYulei2022,mohseni2023}.   An important problem that has remained open so far is to characterize the action of an unknown quantum process on a given set of experimentally accessible input states. In general, such a set will contain more than one  state, and will not be limited to local Clifford invariant states. In intermediate-scale quantum systems, for example, the accessible states generally form  lower-dimensional manifolds of the whole state space, and can be efficiently characterized by low-dimensional representations generated from measurement data~\cite{torlai2018,carrasquilla2019,Ahmed2021PRL,zhu2022}. 
 


Here we develop the first neural network algorithm for characterizing  the action of an unknown quantum
process on a general input state ensemble.
The core of our algorithm is a neural emulator that learns how to mimic the action of the unknown quantum process by operating on an internal representation of the input state ensemble.    The inspiration for our neural emulator comes from the technique of neural style transfer~\cite{karras2019style,karras2020analyzing}, used  in the field of computer graphics to transform images according to a target style.    Similarly to the way an image can be turned into von Gogh's or Picasso's style,   a quantum state can be transformed  according to a target procedure, corresponding to an unknown unitary dynamics or a noisy process.
After being trained on measurement data from a random subset of input/output pairs, the network learns how to transform previously unseen  input states according to the same process,  and to make predictions on the outcome statistics of a  given set of physically relevant measurements.

We test our neural network model on a set of quantum processes in quantum computing, manybody physics, and quantum photonics,   showing that it consistently achieves  high prediction accuracy for parameterized quantum circuits, spin-system nonequivalent dynamics, and continuous-variable non-Gaussian processes over an ensemble of input states, even when trained with  limited numbers of input-output pairs.

{\em Framework.}
The task is to characterize an unknown quantum process, acting on a given quantum system with Hilbert space $\cal H$.  Mathematically, the process can be described by  a quantum channel, that is,  a trace-preserving completely positive linear map $\mathcal E: \mathcal L(\mathcal H)\rightarrow \mathcal L(\mathcal H)$,  where $\mathcal L(\mathcal H)$ denotes the space of linear operators on $\mathcal H$. 
Our goal is to predict the statistics of a  subset  of quantum measurements $\mathcal{M}$ performed on the output of the process $\cal E$,  when the input is drawn from an ensemble $\cal S$ of possible input states. In practice,  $\cal S$ and $\cal M$ can represent two sets of experimentally accessible state preparations and measurements, respectively.   Without loss of generality, we take all measurements in $\cal M$ to have the same number of outcomes, denoted by $m$. 
Mathematically, the measurements in $\cal M$ are described by positive operator-valued measures (POVMs) $\bm{M}  =  (M_k)_{k=1}^{m}$ where each $M_k$ is a positive operator in $L(\cal H)$ and  $\sum_{k=1}^{m} M_k=\mathds{1}$.

\begin{figure}
    \centering
    \includegraphics[width=0.45\textwidth]{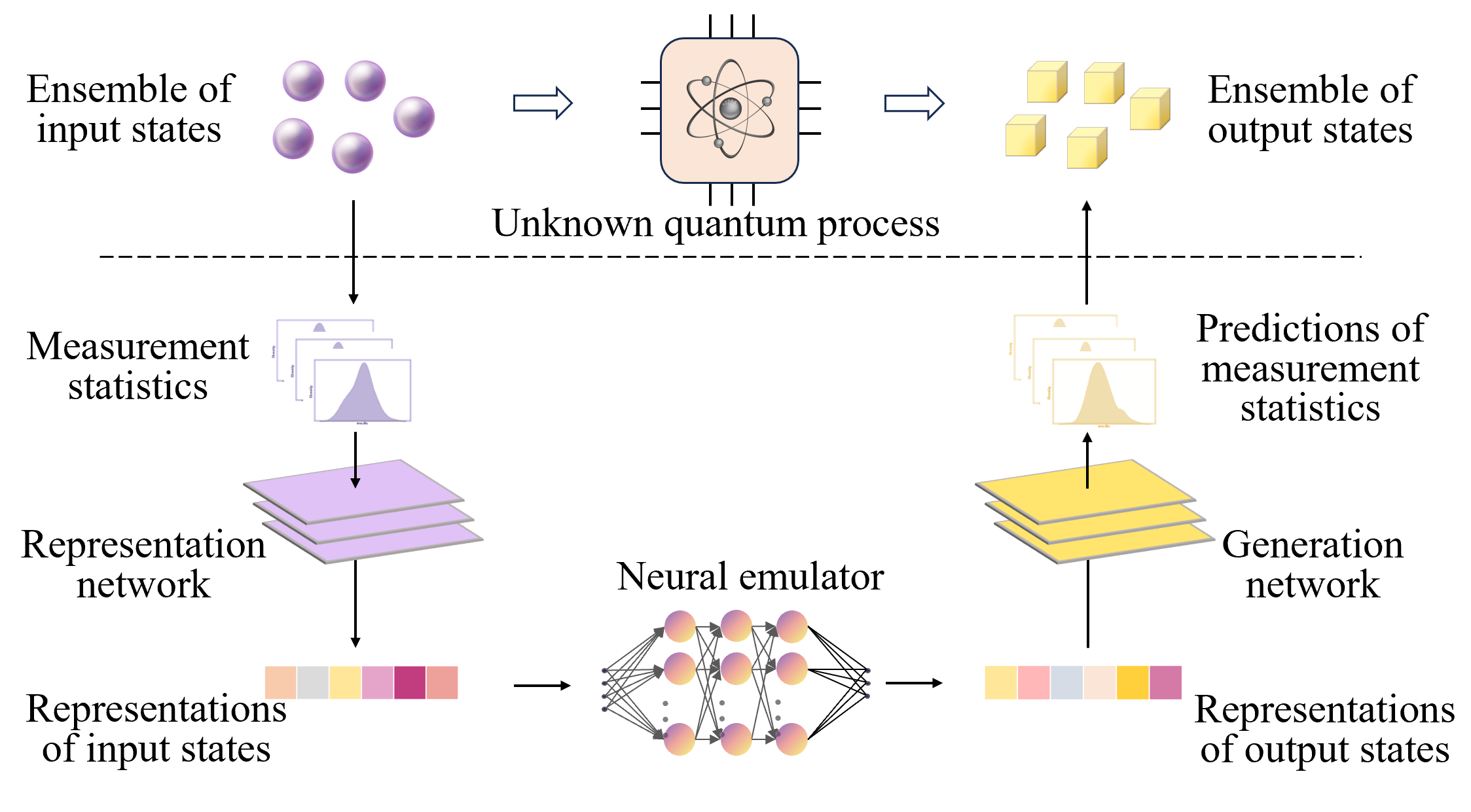}
    \caption{A schematic diagram illustrates how a neural network model, composed of a representation network, a neural emulator and a generation network, learns an unknown quantum process to predict the physical properties of the output.} 
    \label{fig:processLearning}
\end{figure}

 The procedure for characterizing the unknown quantum process $\mathcal E$  consists of two phases: the learning phase and the prediction phase.
 In the learning phase, quantum experiments are used  to acquire  information about the process $\mathcal E$ and to train a neural model to emulate its action.   In  the prediction phase, the experimenter  is asked to predict the  outcome probabilities   $p_k  =  \tr [ M_k  \, {\cal E}  (\rho)]$ associated to a previously unseen state $\rho \in  \cal S$ and to  any measurement $\bm{M}  \in  \cal M$.   We denote the vector of these probabilities by $\bm{p}    =  (p_k)_{k=1}^m$.

The learning phase can be further divided into a measurement subphase and a training subphase. In the measurement subphase, 
the experimenter  collects data by randomly picking some  states in $\cal S$ and all measurements in $\cal M$  and by sampling the corresponding probabilities. 
 In the $i$-th experiment, the experimenter randomly picks  a quantum state $\sigma_i\in \mathcal S$  and prepares many copies of $\sigma_i$. Then the experimenter applies the quantum process $\mathcal E$ on each copy of input state $\sigma_i$, and then performs a measurement  $\bm{M}_j\in\mathcal M$ on the output  state $\mathcal E(\sigma_i)$. By repeating the experiment multiple times, the experimenter then obtains the outcome  statistics $\bm{p}_{ij}^{({\rm out})}$.  In the typical situation, we assume that the number of repetitions of the experiment is sufficiently large that the statistics $\bm{p}_{ij}^{({\rm out})}$ is close to the  ideal probability distribution $\tr(\mathcal E(\sigma_i)\bm{M}_j)$. 
It is important to stress  that neither the set $\mathcal S$ of input states nor the set $\mathcal M$ of output measurements need to be information complete.
In fact, the experimenter may not even be able to  actively choose the input states and the output measurements, as long as both the states and the measurements are  
sampled from probability distributions that do not change too much from the learning phase to the prediction phase. 

In the training subphase, the data are used to establish a  predictive model for the quantum process $\mathcal E$. This task is achieved by the neural  model illustrated in Figure \ref{fig:processLearning}.  
   The  model is a combination of a representation network, a neural emulator and a generation network.  The representation network  uses  measurement data to construct representations of quantum states, while the generation network uses the state representations to predict the statistics of measurements performed on the corresponding states.    The neural emulator learns a mapping between input state representations and output state representations, which can be regarded as a low-dimensional analogue of the quantum process~$\mathcal E$.

   The combination of the representation network and the generation network is known as a generative query neural network for quantum states (GQNQ)~\cite{zhu2022}. To train our model, we first train GQNQ and then the neural emulator by minimizing differences between predictions  and ground truths.   The model   is provided  the measurement data $\mathcal D:=\{(\bm{m}_j,  \bm{p}_{ij}^{({\rm in })}, \bm{p}_{ij}^{({\rm out})}): 1\le i\le n, 1\le j\le |\mathcal M|\}$, where  $\bm{p}_{ij}^{({\rm in})}:=\tr(\sigma_i\bm{M}_j)$ is the  statistics of measurement $\bm{M}_j$ on input state $\sigma_i$, and $\bm{m}_j$ is either the full description  of the measurement  operators in $\bm{M}_j$ or a low-dimensional parametrization, specifically valid for the set  $\cal M$ of physically accessible measurements.  

In the prediction phase, the experimenter no longer has access to the  quantum process $\mathcal E$. 
 The goal is to predict the measurement statistics on the output  of the process, when the input is a new, possibly unknown quantum state $\rho\in\mathcal S$.  The experimenter can access a number of copies of the state, thereby estimating the statistics for all of the measurements  in the physically accessible set $\cal M$.  Based on the resulting data, the experimenter then predicts 
the measurement statistics $\left(\tr(\mathcal E(\rho) M^{(i)})\right)_{i=1}^m$ for $\forall \bm{M}\in \mathcal M$ on the output state $\mathcal E(\rho)$. 
  When both the set of states $\mathcal S$  and the set of measurements $\mathcal M$ are information complete, this task is reduced to quantum process tomography. For large quantum systems, instead, the sets of physically accessible states and measurements  generally form  lower-dimensional manifolds, and the output statistics  can be accurately predicted without a full process tomography. Leveraging on this fact, our  model can make high-quality predictions even with limited data, as shown in the following applications.

\begin{figure}[hbtp]
    \includegraphics[width=0.42\textwidth]{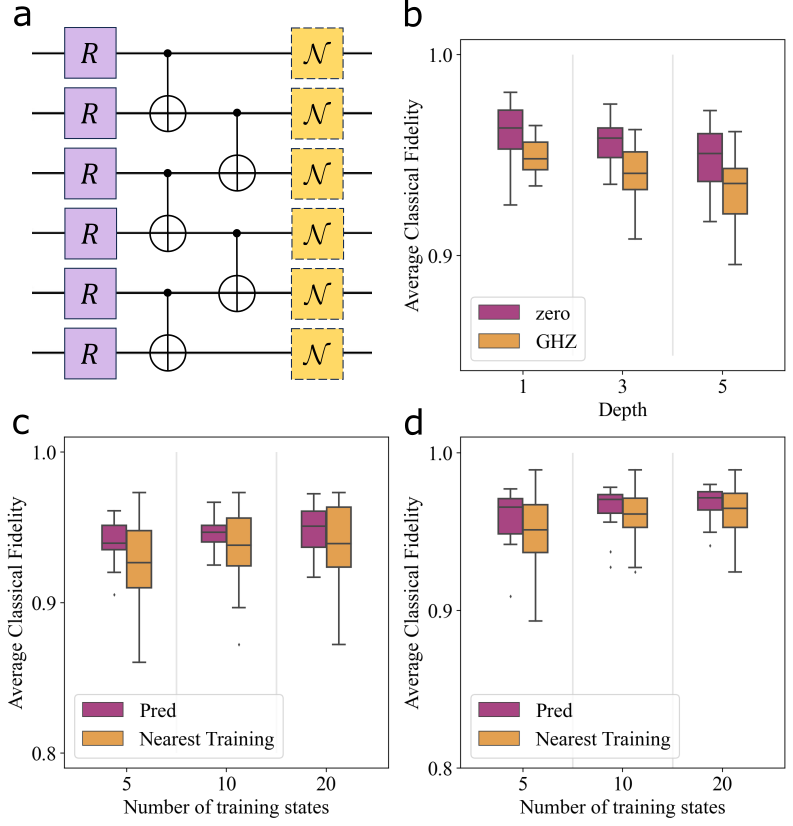}
    \caption{ The numerical results of learning a six-qubit parametrized quantum circuit, visualized by box plots~\cite{krzywinski2014visualizing}.
    a.\ Structure of each layer of the quantum circuit under consideration.
    b.\ The average classical fidelity between prediction of outcome statistic and the ground truths versus the depth of a quantum circuit for two different types of input states. c.\ Comparison between the prediction of measurement statistics given by the neural network and that of the nearest training data for a three-layer unitary quantum circuit using different number of sampled training input-output pairs. d.\ Comparison between the prediction given by the neural network and that of the nearest training data for a noisy quantum circuit using different number of training states.}
    \label{fig:circuitLearning}
\end{figure}

{\em Learning parametrized quantum circuits.}
We start by using our neural network model to learn a six-qubit parametrized quantum circuit,  containing layers of single-qubit rotations followed by layers of two-qubit controlled-NOT gates on all nearest-neighbour pair of qubits, as shown in Fig.~\ref{fig:circuitLearning}a. We require our neural network to predict  the outcome statistics of all six-qubit Pauli measurements performed on the output of a quantum circuit with the parameters of each quantum gate chosen uniformly at random.  We consider two types of input quantum states, including locally rotated zero state $\otimes_{i=1}^6  (U_i\ket{0})^{\otimes 6}$ and locally rotated GHZ state $1/\sqrt{2}\otimes_{i=1}^6 U_i (\ket{000000}+\ket{111111})$, where $ U_i= \exp(-\textrm{i}\theta_{i,z} \sigma_{i,z}) \exp(-\textrm{i}\theta_{i,y} \sigma_{i,y}) \exp(-\textrm{i}\theta_{i,x}\sigma_{i,x})$ is a single-qubit unitary rotation and $\theta_{i,x}, \theta_{i,y}, \theta_{i,z}\in [0, 0.3\pi]$. Fig.~\ref{fig:circuitLearning}b shows the average classical fidelity between predictions and calculated truths, averaged over $20$ test states and all the possible Pauli measurements, versus the depth of the quantum circuit, where the predictions are made by a neural network trained over $n=20$ input-output pairs.  As the circuit depth increases, the average classical fidelity decreases. Numerical results suggest that predicting the output corresponding to highly entangled input is harder than that corresponding to product state input.

To benchmark the performance of our predictive model, it is helpful to compare the prediction made by our model with a simple baseline strategy known as the nearest training data strategy~\cite{huang2022provably}. For a given test state $\rho$ and query measurement $\bm M$, the strategy consists in searching for the  training input state $\sigma$ that is closest to $\rho$ in terms of quantum fidelity and outputting  the probability distribution $\bm{p}_{\sigma, \bm{m}}^{({\rm out})}$ in the dataset $\mathcal D$ as the prediction for $\tr(\rho \bm{M})$. Fig.~\ref{fig:circuitLearning}c shows the classical fidelities (averaged over all the possible Pauli measurements) of the predictions given by both approaches over $20$ test states with respect to different number of training input-output state pairs. We find that the average accuracy of our neural network model is higher than that of the nearest training data strategy, and the variance of the performance of our neural network is lower than that of the nearest training data strategy.

Furthermore, we apply our neural network model for learning a three-layer noisy quantum circuit, where each qubit in each layer suffers a depolarizing noise with error rate $0.01$ as shown in Fig.~\ref{fig:circuitLearning}a. 
Fig.~\ref{fig:circuitLearning}d compares the average classical fidelities of the prediction made by our neural network model and that given by the nearest training data strategy for different numbers of training input-output state pairs $n=5, 10$, and $20$, where again, the classical fidelity is averaged over $20$ test states and all the possible Pauli measurements.

{\em Learning nonequilibrium quantum dynamics.}
A natural testbed for our neural network model is nonequilibrium quantum dynamics of spin systems.
Here we first apply our neural network model for learning a quantum dynamic initialized in a state $\ket{+}^{\otimes L}$ with local rotations governed by 
a Hamiltonian in long-range Ising model with transverse field, which has been experimentally investigated in Refs.~\cite{lanyon2017,friis2018}.
The initial quantum state at time $t=0$ is uniformly randomly selected from the ensemble of product states $\left\{ \otimes_{i=1}^L U_i \ket{+}^{\otimes 6}: \theta_{i,z}, \theta_{i,y}, \theta_{i,x}\in [0, \pi/10] \right\}$. 
Then it undergoes a quantum unitary evolution $U(t):=e^{-\textit{i}H t}$ with Hamiltonian
$H=-\sum_{0\le i<j \le L-1} J_{ij}\sigma_i^x\sigma_{j}^x -B\sum_{j=0}^{L-1} \sigma_j^z$,
where $J_{ij}:=1/|i-j|^\alpha$ denotes the interaction strength between $i$th and $j$th spins, following a power law decay.

\begin{figure}
    \centering
    \includegraphics[width=0.35\textwidth]{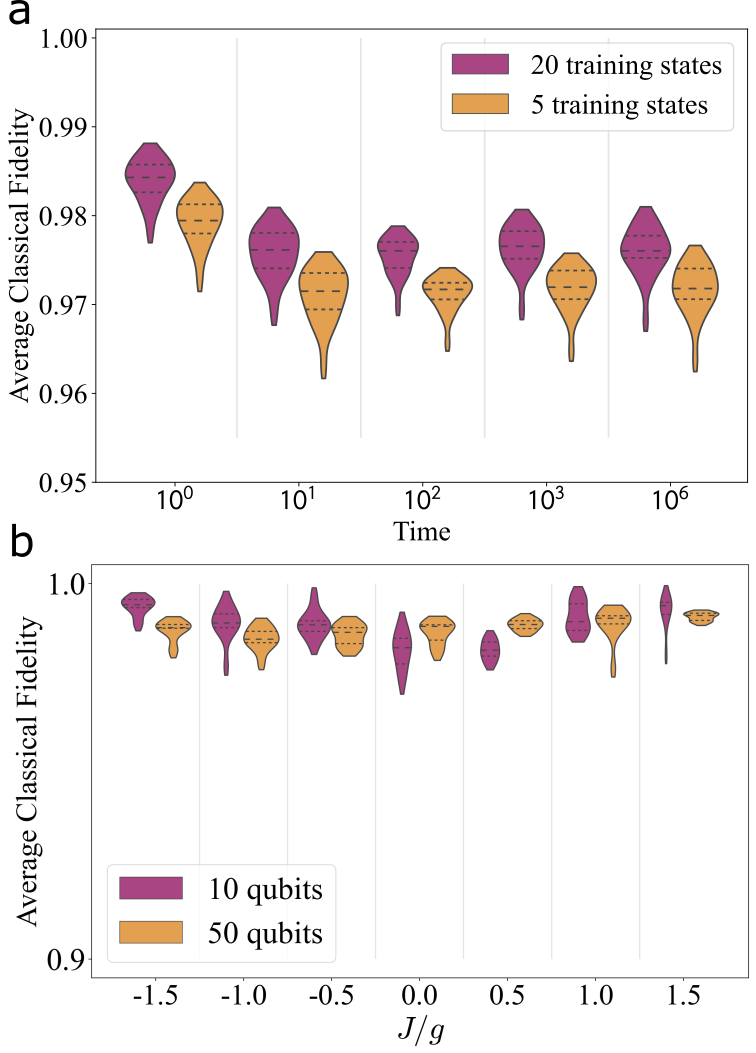}
    \caption{The numerical results of learning spin-system quantum quantum processes, visualized by violin plots~\cite{hintze1998violin}. 
    a.\ The average classical fidelity of the predictions of the measurement statistics on the final state at different time intervals. b.\ The average classical fidelity of the predictions for the quench dynamics driven by Ising model versus different ratios of $J/g$. }
    \label{fig:dynamics}
\end{figure}

We consider six-qubit ($L=6$) quantum dynamics with $\alpha=0.01$ and choose $\mathcal M$ as the set of all six-qubit Pauli measurements.
It is interesting to see how the evolution time affects the prediction performance of our neural network model in this case.
In Fig.\ref{fig:dynamics}a, we show the average classical fidelity between the predictions and the calculated truths for various time intervals of unitary evolution: $t=1, 10, 100, 10^3$, and $10^6$, where the predictions for each evolution time are made by an independent neural network model over $20$ test states. 
The results indicate that the predictions made by our neural network remains high accuracy even for significantly extended evolution time, providing a shortcut for the emulation of a long-time dynamical quantum process.

Next we apply our neural network model for learning an intermediate scale quantum quench dynamics.
The unitary quantum process we consider is $U=\exp(-\textit{i}t H_{\text{Ising}})$, driven by the Ising Hamiltonian 
$H_{\text{Ising}}=J\sum_{i=0}^{L-2} \sigma_i^z\sigma_{i+1}^z+\sum_{j=0}^{L-1}  \sigma_j^x$,
with $J=0.5$ and $t=10$. 
The ensemble $\mathcal S$ of initial states are the ground states of site-dependent
Ising model described by 
$H_{\text{sdIsing}}=\sum_{i=0}^{L-2} J_i\sigma_i^z\sigma_{i+1}^z+\sum_{j=0}^{L-1}  \sigma_j^x$,
with each $J_i$ being an independent Gaussian random variable $\mathcal N(J, 1)$.
The set $\mathcal M$ of measurements consists of all the possible three-local Pauli measurements. 

We train our neural network with five different initial ground states for each value of $J\in \{-1.5, -1, \cdots, 1.5\}$.
After the training is concluded, we test our trained model for another 10 different initial ground states for each value of $J$ and the classical fidelity between the predictions made by our neural network model and the calculated truths are shown in Fig.~\ref{fig:dynamics}b for both $L=10$ and $L=50$.
The prediction in the parameter region where $J$ is far from zero is more accurate than that when $J=0$ since the initial states become more complex due to the equal probability of ferromagnetic and antiferromagnetic interactions when $J=0$.

{\em Learning nonlinear optical quantum processes.}
Our neural network model can also be used to learn continuous-variable nonGaussian quantum processes. Here we consider an important quantum process, induced by Kerr nonlinear interaction $K= \hat{a}^{\dagger\, 2}\hat{a}^2$~\cite{dykman2012}, where $\hat{a}$ and $\hat{a}^\dagger$ are the annihilation and creation operators respectively.  The quantum process we consider is a unitary transformation $U=\text{e}^{-\textrm{i}K t}$, where $t=1$. The set $\mathcal S$ of input states are composed of coherent states $\ket{\alpha}$ with amplitude $\alpha=r \textrm{e}^{\textrm{i}\psi}$, where $r\in [0,3]$ and $\psi\in [0,2\pi)$ are both uniformly randomly chosen. The measurement set  $\mathcal M$ is a set of homodyne measurements, i.e.\ the projective measurements associated to quadrature operators $(\textrm{e}^{\textrm{i}\theta} \hat{a}^\dagger +\textrm{e}^{-\textrm{i}\theta}\hat{a})/2$ with phase $\theta\in \{0, \pi/100, \dots, 99\pi/100\}$.

\begin{figure}[hbtp]
    \centering
    \includegraphics[width=0.35\textwidth]{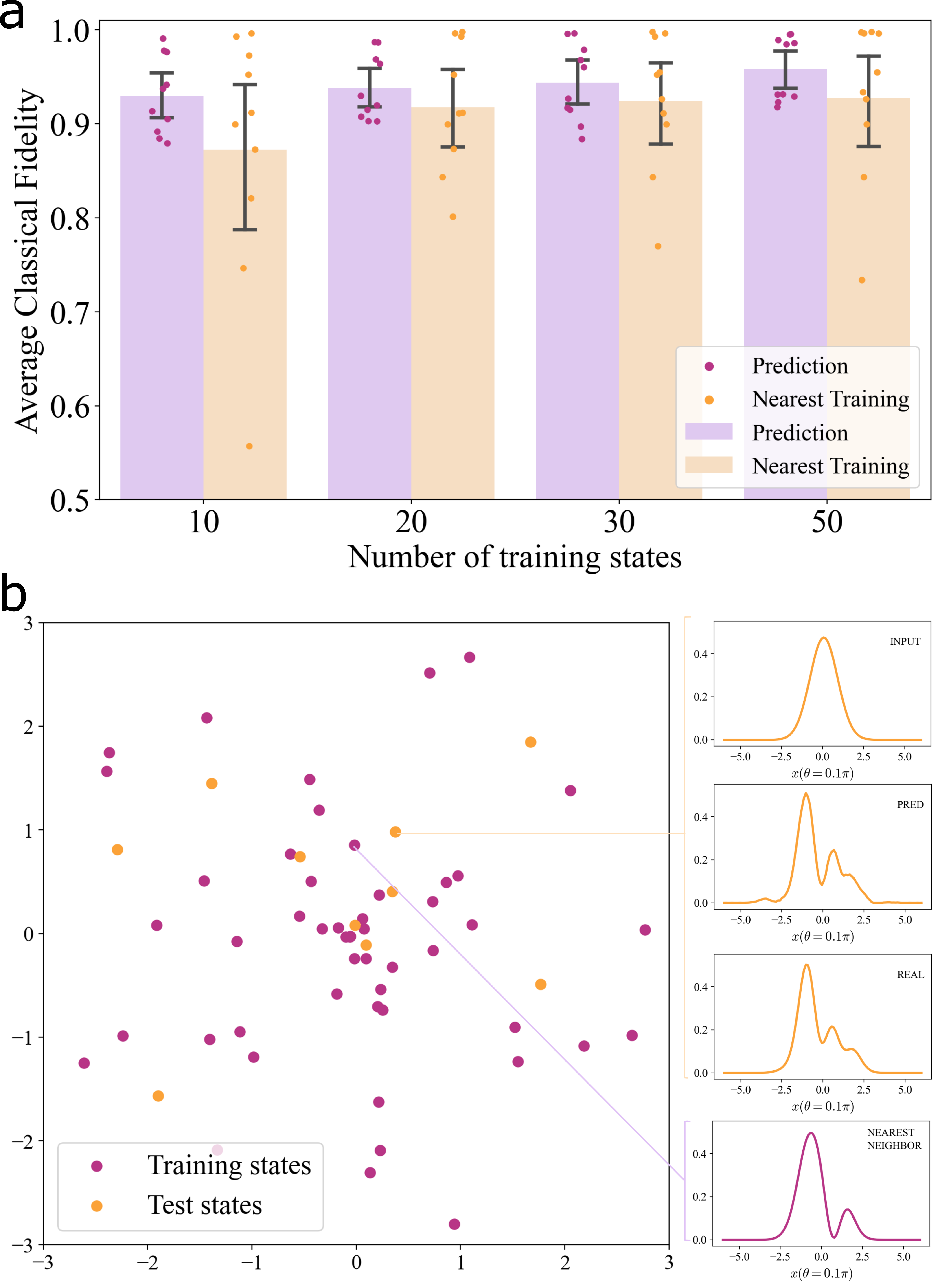}
    \caption{The numerical results of learning a continuous-variable quantum process. a.\ Comparison between the performance of the neural network model and nearest training data strategy.
    Each violet (orange) bar shows the average classical fidelity between the prediction produced by the neural network model (nearest training data) and the ground truths. Each dot corresponds to one test input. b.\ the amplitudes of coherent states for both training and test are shown, together with an example of the true outcome distribution at an input and the corresponding output, as well as the prediction of outcome distribution produced by our neural network and the nearest training data strategy. 
    }
    \label{fig:continuousResults}
\end{figure}

In the learning phase, we randomly sample $n$ coherent states from the ensemble as the input states, whose amplitudes are shown by violet dots in Fig.~\ref{fig:continuousResults}b, for instance, when $n=50$, injected into the quantum process and use the measurement statistics of the corresponding output states for training our neural network model. In the prediction phase, we randomly sample another 10 different coherent states from the ensemble,  whose amplitudes are shown by orange dots in Fig.~\ref{fig:continuousResults}b, and test the performance of the trained model for these ten test input states. Fig.~\ref{fig:continuousResults}a shows the average classical fidelity between the predicted outcome statistics produced by our neural network model and the calculated ground truths averaged over all the test states and all the possible quadrature bases, when the number of training states are $n=10, 20, 30$ and $50$.

The average classical fidelity of the prediction generated by our predictive model is clearly higher than that of the nearest training data strategy.
In Fig.~\ref{fig:continuousResults}b, we show the amplitudes of both the training coherent states and test coherent states on the phase space.  We also provide an example demonstrating the neural network's prediction of a specific measurement outcome distribution, as well as a comparison with the nearest training data, together with the calculated true outcome distribution.
The shape of the outcome probability distribution produced by our neural model is much closer to the true distribution in this example. These results indicate that our neural network model's predictive power about this Kerr gate is well generalized to those coherent states unseen during the training phase.

{\em Conclusions.} 
We have developed a neural network model for characterizing the action of unknown quantum processes applied on a given ensemble of input states.
Our approach can be used in realistic scenarios where the experimenter has limited quantum technology, and therefore has access only to limited experimental data.     During the learning phase, the experimenter  only needs to prepare quantum states in an ensemble of accessible states and to perform quantum measurements in a  set of accessible measurements. In the prediction phase, only classical information processing is required.   These features make our results appealing for practical applications,  such as  learning a noisy quantum circuit for error mitigation~\cite{temme2017,endo2018}, predicting the evolution outcome of quantum dynamics~\cite{flurin2020,mohseni2022,koolstra2022,huangYulei2022,mohseni2023}, and benchmarking optical quantum devices~\cite{adesso2008,chiribella2013}.


{\em Acknowledgements.}
This work was supported by funding from the Hong Kong Research Grant Council through grants no.\ 17300918 and no.\ 17307520, through the Senior Research Fellowship Scheme SRFS2021-7S02, and through the Early Career Scheme (ECS) grant 27310822, the Croucher Foundation, the John Templeton Foundation through grant  62312, The Quantum Information Structure of Spacetime (qiss.fr), and by Guangdong Basic and Applied Basic Research Foundation (Project No.~2022A1515010340).
YXW acknowledges funding from the National Natural Science Foundation of China through grants no.\ 61872318. Research at the Perimeter Institute is supported by the Government of Canada through the Department of Innovation, Science and Economic Development Canada and by the Province of Ontario through the Ministry of Research, Innovation and Science. The opinions expressed in this publication are those of the authors and do not necessarily reflect the views of the John Templeton Foundation.

\bibliography{refs}

\begin{thebibliography}{45}%
\makeatletter
\providecommand \@ifxundefined [1]{%
 \@ifx{#1\undefined}
}%
\providecommand \@ifnum [1]{%
 \ifnum #1\expandafter \@firstoftwo
 \else \expandafter \@secondoftwo
 \fi
}%
\providecommand \@ifx [1]{%
 \ifx #1\expandafter \@firstoftwo
 \else \expandafter \@secondoftwo
 \fi
}%
\providecommand \natexlab [1]{#1}%
\providecommand \enquote  [1]{``#1''}%
\providecommand \bibnamefont  [1]{#1}%
\providecommand \bibfnamefont [1]{#1}%
\providecommand \citenamefont [1]{#1}%
\providecommand \href@noop [0]{\@secondoftwo}%
\providecommand \href [0]{\begingroup \@sanitize@url \@href}%
\providecommand \@href[1]{\@@startlink{#1}\@@href}%
\providecommand \@@href[1]{\endgroup#1\@@endlink}%
\providecommand \@sanitize@url [0]{\catcode `\\12\catcode `\$12\catcode
  `\&12\catcode `\#12\catcode `\^12\catcode `\_12\catcode `\%12\relax}%
\providecommand \@@startlink[1]{}%
\providecommand \@@endlink[0]{}%
\providecommand \url  [0]{\begingroup\@sanitize@url \@url }%
\providecommand \@url [1]{\endgroup\@href {#1}{\urlprefix }}%
\providecommand \urlprefix  [0]{URL }%
\providecommand \Eprint [0]{\href }%
\providecommand \doibase [0]{http://dx.doi.org/}%
\providecommand \selectlanguage [0]{\@gobble}%
\providecommand \bibinfo  [0]{\@secondoftwo}%
\providecommand \bibfield  [0]{\@secondoftwo}%
\providecommand \translation [1]{[#1]}%
\providecommand \BibitemOpen [0]{}%
\providecommand \bibitemStop [0]{}%
\providecommand \bibitemNoStop [0]{.\EOS\space}%
\providecommand \EOS [0]{\spacefactor3000\relax}%
\providecommand \BibitemShut  [1]{\csname bibitem#1\endcsname}%
\let\auto@bib@innerbib\@empty
\bibitem [{\citenamefont {Chuang}\ and\ \citenamefont
  {Nielsen}(1997)}]{chuang1997}%
  \BibitemOpen
  \bibfield  {author} {\bibinfo {author} {\bibfnamefont {Isaac~L}\ \bibnamefont
  {Chuang}}\ and\ \bibinfo {author} {\bibfnamefont {Michael~A}\ \bibnamefont
  {Nielsen}},\ }\bibfield  {title} {\enquote {\bibinfo {title} {Prescription
  for experimental determination of the dynamics of a quantum black box},}\
  }\href@noop {} {\bibfield  {journal} {\bibinfo  {journal} {J. Mod. Opt.}\
  }\textbf {\bibinfo {volume} {44}},\ \bibinfo {pages} {2455--2467} (\bibinfo
  {year} {1997})}\BibitemShut {NoStop}%
\bibitem [{\citenamefont {D'Ariano}\ and\ \citenamefont
  {Lo~Presti}(2001)}]{dariano2001}%
  \BibitemOpen
  \bibfield  {author} {\bibinfo {author} {\bibfnamefont {G.~M.}\ \bibnamefont
  {D'Ariano}}\ and\ \bibinfo {author} {\bibfnamefont {P.}~\bibnamefont
  {Lo~Presti}},\ }\bibfield  {title} {\enquote {\bibinfo {title} {Quantum
  tomography for measuring experimentally the matrix elements of an arbitrary
  quantum operation},}\ }\href {\doibase 10.1103/PhysRevLett.86.4195}
  {\bibfield  {journal} {\bibinfo  {journal} {Phys. Rev. Lett.}\ }\textbf
  {\bibinfo {volume} {86}},\ \bibinfo {pages} {4195--4198} (\bibinfo {year}
  {2001})}\BibitemShut {NoStop}%
\bibitem [{\citenamefont {Altepeter}\ \emph {et~al.}(2003)\citenamefont
  {Altepeter}, \citenamefont {Branning}, \citenamefont {Jeffrey}, \citenamefont
  {Wei}, \citenamefont {Kwiat}, \citenamefont {Thew}, \citenamefont {O'Brien},
  \citenamefont {Nielsen},\ and\ \citenamefont {White}}]{altepeter2003}%
  \BibitemOpen
  \bibfield  {author} {\bibinfo {author} {\bibfnamefont {J.~B.}\ \bibnamefont
  {Altepeter}}, \bibinfo {author} {\bibfnamefont {D.}~\bibnamefont {Branning}},
  \bibinfo {author} {\bibfnamefont {E.}~\bibnamefont {Jeffrey}}, \bibinfo
  {author} {\bibfnamefont {T.~C.}\ \bibnamefont {Wei}}, \bibinfo {author}
  {\bibfnamefont {P.~G.}\ \bibnamefont {Kwiat}}, \bibinfo {author}
  {\bibfnamefont {R.~T.}\ \bibnamefont {Thew}}, \bibinfo {author}
  {\bibfnamefont {J.~L.}\ \bibnamefont {O'Brien}}, \bibinfo {author}
  {\bibfnamefont {M.~A.}\ \bibnamefont {Nielsen}}, \ and\ \bibinfo {author}
  {\bibfnamefont {A.~G.}\ \bibnamefont {White}},\ }\bibfield  {title} {\enquote
  {\bibinfo {title} {Ancilla-assisted quantum process tomography},}\ }\href
  {\doibase 10.1103/PhysRevLett.90.193601} {\bibfield  {journal} {\bibinfo
  {journal} {Phys. Rev. Lett.}\ }\textbf {\bibinfo {volume} {90}},\ \bibinfo
  {pages} {193601} (\bibinfo {year} {2003})}\BibitemShut {NoStop}%
\bibitem [{\citenamefont {Lobino}\ \emph {et~al.}(2008)\citenamefont {Lobino},
  \citenamefont {Korystov}, \citenamefont {Kupchak}, \citenamefont {Figueroa},
  \citenamefont {Sanders},\ and\ \citenamefont {Lvovsky}}]{lobino2008}%
  \BibitemOpen
  \bibfield  {author} {\bibinfo {author} {\bibfnamefont {Mirko}\ \bibnamefont
  {Lobino}}, \bibinfo {author} {\bibfnamefont {Dmitry}\ \bibnamefont
  {Korystov}}, \bibinfo {author} {\bibfnamefont {Connor}\ \bibnamefont
  {Kupchak}}, \bibinfo {author} {\bibfnamefont {Eden}\ \bibnamefont
  {Figueroa}}, \bibinfo {author} {\bibfnamefont {Barry~C}\ \bibnamefont
  {Sanders}}, \ and\ \bibinfo {author} {\bibfnamefont {AI}~\bibnamefont
  {Lvovsky}},\ }\bibfield  {title} {\enquote {\bibinfo {title} {Complete
  characterization of quantum-optical processes},}\ }\href@noop {} {\bibfield
  {journal} {\bibinfo  {journal} {Science}\ }\textbf {\bibinfo {volume}
  {322}},\ \bibinfo {pages} {563--566} (\bibinfo {year} {2008})}\BibitemShut
  {NoStop}%
\bibitem [{\citenamefont {Rahimi-Keshari}\ \emph {et~al.}(2011)\citenamefont
  {Rahimi-Keshari}, \citenamefont {Scherer}, \citenamefont {Mann},
  \citenamefont {Rezakhani}, \citenamefont {Lvovsky},\ and\ \citenamefont
  {Sanders}}]{rahimi2011}%
  \BibitemOpen
  \bibfield  {author} {\bibinfo {author} {\bibfnamefont {Saleh}\ \bibnamefont
  {Rahimi-Keshari}}, \bibinfo {author} {\bibfnamefont {Artur}\ \bibnamefont
  {Scherer}}, \bibinfo {author} {\bibfnamefont {Ady}\ \bibnamefont {Mann}},
  \bibinfo {author} {\bibfnamefont {Ali~T}\ \bibnamefont {Rezakhani}}, \bibinfo
  {author} {\bibfnamefont {AI}~\bibnamefont {Lvovsky}}, \ and\ \bibinfo
  {author} {\bibfnamefont {Barry~C}\ \bibnamefont {Sanders}},\ }\bibfield
  {title} {\enquote {\bibinfo {title} {Quantum process tomography with coherent
  states},}\ }\href@noop {} {\bibfield  {journal} {\bibinfo  {journal} {New J.
  Phys.}\ }\textbf {\bibinfo {volume} {13}},\ \bibinfo {pages} {013006}
  (\bibinfo {year} {2011})}\BibitemShut {NoStop}%
\bibitem [{\citenamefont {Aaronson}(2018)}]{aaronson2018}%
  \BibitemOpen
  \bibfield  {author} {\bibinfo {author} {\bibfnamefont {Scott}\ \bibnamefont
  {Aaronson}},\ }\bibfield  {title} {\enquote {\bibinfo {title} {Shadow
  tomography of quantum states},}\ }in\ \href@noop {} {\emph {\bibinfo
  {booktitle} {Proceedings of the 50th Annual ACM SIGACT Symposium on Theory of
  Computing}}}\ (\bibinfo {year} {2018})\ pp.\ \bibinfo {pages}
  {325--338}\BibitemShut {NoStop}%
\bibitem [{\citenamefont {Huang}\ \emph {et~al.}(2020)\citenamefont {Huang},
  \citenamefont {Kueng},\ and\ \citenamefont {Preskill}}]{huang2020}%
  \BibitemOpen
  \bibfield  {author} {\bibinfo {author} {\bibfnamefont {Hsin-Yuan}\
  \bibnamefont {Huang}}, \bibinfo {author} {\bibfnamefont {Richard}\
  \bibnamefont {Kueng}}, \ and\ \bibinfo {author} {\bibfnamefont {John}\
  \bibnamefont {Preskill}},\ }\bibfield  {title} {\enquote {\bibinfo {title}
  {Predicting many properties of a quantum system from very few
  measurements},}\ }\href@noop {} {\bibfield  {journal} {\bibinfo  {journal}
  {Nat. Phys.}\ }\textbf {\bibinfo {volume} {16}},\ \bibinfo {pages}
  {1050--1057} (\bibinfo {year} {2020})}\BibitemShut {NoStop}%
\bibitem [{\citenamefont {Huang}\ \emph
  {et~al.}(2022{\natexlab{a}})\citenamefont {Huang}, \citenamefont {Chen},\
  and\ \citenamefont {Preskill}}]{huang2022}%
  \BibitemOpen
  \bibfield  {author} {\bibinfo {author} {\bibfnamefont {Hsin-Yuan}\
  \bibnamefont {Huang}}, \bibinfo {author} {\bibfnamefont {Sitan}\ \bibnamefont
  {Chen}}, \ and\ \bibinfo {author} {\bibfnamefont {John}\ \bibnamefont
  {Preskill}},\ }\bibfield  {title} {\enquote {\bibinfo {title} {Learning to
  predict arbitrary quantum processes},}\ }\href@noop {} {\bibfield  {journal}
  {\bibinfo  {journal} {arXiv:2210.14894}\ } (\bibinfo {year}
  {2022}{\natexlab{a}})}\BibitemShut {NoStop}%
\bibitem [{\citenamefont {Flurin}\ \emph {et~al.}(2020)\citenamefont {Flurin},
  \citenamefont {Martin}, \citenamefont {Hacohen-Gourgy},\ and\ \citenamefont
  {Siddiqi}}]{flurin2020}%
  \BibitemOpen
  \bibfield  {author} {\bibinfo {author} {\bibfnamefont {E.}~\bibnamefont
  {Flurin}}, \bibinfo {author} {\bibfnamefont {L.~S.}\ \bibnamefont {Martin}},
  \bibinfo {author} {\bibfnamefont {S.}~\bibnamefont {Hacohen-Gourgy}}, \ and\
  \bibinfo {author} {\bibfnamefont {I.}~\bibnamefont {Siddiqi}},\ }\bibfield
  {title} {\enquote {\bibinfo {title} {Using a recurrent neural network to
  reconstruct quantum dynamics of a superconducting qubit from physical
  observations},}\ }\href {\doibase 10.1103/PhysRevX.10.011006} {\bibfield
  {journal} {\bibinfo  {journal} {Phys. Rev. X}\ }\textbf {\bibinfo {volume}
  {10}},\ \bibinfo {pages} {011006} (\bibinfo {year} {2020})}\BibitemShut
  {NoStop}%
\bibitem [{\citenamefont {Mohseni}\ \emph {et~al.}(2022)\citenamefont
  {Mohseni}, \citenamefont {F{\"o}sel}, \citenamefont {Guo}, \citenamefont
  {Navarrete-Benlloch},\ and\ \citenamefont {Marquardt}}]{mohseni2022}%
  \BibitemOpen
  \bibfield  {author} {\bibinfo {author} {\bibfnamefont {Naeimeh}\ \bibnamefont
  {Mohseni}}, \bibinfo {author} {\bibfnamefont {Thomas}\ \bibnamefont
  {F{\"o}sel}}, \bibinfo {author} {\bibfnamefont {Lingzhen}\ \bibnamefont
  {Guo}}, \bibinfo {author} {\bibfnamefont {Carlos}\ \bibnamefont
  {Navarrete-Benlloch}}, \ and\ \bibinfo {author} {\bibfnamefont {Florian}\
  \bibnamefont {Marquardt}},\ }\bibfield  {title} {\enquote {\bibinfo {title}
  {Deep learning of quantum many-body dynamics via random driving},}\
  }\href@noop {} {\bibfield  {journal} {\bibinfo  {journal} {Quantum}\ }\textbf
  {\bibinfo {volume} {6}},\ \bibinfo {pages} {714} (\bibinfo {year}
  {2022})}\BibitemShut {NoStop}%
\bibitem [{\citenamefont {Koolstra}\ \emph {et~al.}(2022)\citenamefont
  {Koolstra}, \citenamefont {Stevenson}, \citenamefont {Barzili}, \citenamefont
  {Burns}, \citenamefont {Siva}, \citenamefont {Greenfield}, \citenamefont
  {Livingston}, \citenamefont {Hashim}, \citenamefont {Naik}, \citenamefont
  {Kreikebaum}, \citenamefont {O'Brien}, \citenamefont {Santiago},
  \citenamefont {Dressel},\ and\ \citenamefont {Siddiqi}}]{koolstra2022}%
  \BibitemOpen
  \bibfield  {author} {\bibinfo {author} {\bibfnamefont {G.}~\bibnamefont
  {Koolstra}}, \bibinfo {author} {\bibfnamefont {N.}~\bibnamefont {Stevenson}},
  \bibinfo {author} {\bibfnamefont {S.}~\bibnamefont {Barzili}}, \bibinfo
  {author} {\bibfnamefont {L.}~\bibnamefont {Burns}}, \bibinfo {author}
  {\bibfnamefont {K.}~\bibnamefont {Siva}}, \bibinfo {author} {\bibfnamefont
  {S.}~\bibnamefont {Greenfield}}, \bibinfo {author} {\bibfnamefont
  {W.}~\bibnamefont {Livingston}}, \bibinfo {author} {\bibfnamefont
  {A.}~\bibnamefont {Hashim}}, \bibinfo {author} {\bibfnamefont {R.~K.}\
  \bibnamefont {Naik}}, \bibinfo {author} {\bibfnamefont {J.~M.}\ \bibnamefont
  {Kreikebaum}}, \bibinfo {author} {\bibfnamefont {K.~P.}\ \bibnamefont
  {O'Brien}}, \bibinfo {author} {\bibfnamefont {D.~I.}\ \bibnamefont
  {Santiago}}, \bibinfo {author} {\bibfnamefont {J.}~\bibnamefont {Dressel}}, \
  and\ \bibinfo {author} {\bibfnamefont {I.}~\bibnamefont {Siddiqi}},\
  }\bibfield  {title} {\enquote {\bibinfo {title} {Monitoring fast
  superconducting qubit dynamics using a neural network},}\ }\href {\doibase
  10.1103/PhysRevX.12.031017} {\bibfield  {journal} {\bibinfo  {journal} {Phys.
  Rev. X}\ }\textbf {\bibinfo {volume} {12}},\ \bibinfo {pages} {031017}
  (\bibinfo {year} {2022})}\BibitemShut {NoStop}%
\bibitem [{\citenamefont {Huang}\ \emph
  {et~al.}(2022{\natexlab{b}})\citenamefont {Huang}, \citenamefont {Che},
  \citenamefont {Wei}, \citenamefont {Xu}, \citenamefont {Nie}, \citenamefont
  {Li}, \citenamefont {Lu},\ and\ \citenamefont {Xin}}]{huangYulei2022}%
  \BibitemOpen
  \bibfield  {author} {\bibinfo {author} {\bibfnamefont {Yulei}\ \bibnamefont
  {Huang}}, \bibinfo {author} {\bibfnamefont {Liangyu}\ \bibnamefont {Che}},
  \bibinfo {author} {\bibfnamefont {Chao}\ \bibnamefont {Wei}}, \bibinfo
  {author} {\bibfnamefont {Feng}\ \bibnamefont {Xu}}, \bibinfo {author}
  {\bibfnamefont {Xinfang}\ \bibnamefont {Nie}}, \bibinfo {author}
  {\bibfnamefont {Jun}\ \bibnamefont {Li}}, \bibinfo {author} {\bibfnamefont
  {Dawei}\ \bibnamefont {Lu}}, \ and\ \bibinfo {author} {\bibfnamefont {Tao}\
  \bibnamefont {Xin}},\ }\bibfield  {title} {\enquote {\bibinfo {title}
  {Measuring quantum entanglement from local information by machine
  learning},}\ }\href@noop {} {\bibfield  {journal} {\bibinfo  {journal}
  {arXiv:2209.08501}\ } (\bibinfo {year} {2022}{\natexlab{b}})}\BibitemShut
  {NoStop}%
\bibitem [{\citenamefont {Mohseni}\ \emph {et~al.}(2023)\citenamefont
  {Mohseni}, \citenamefont {Shi}, \citenamefont {Byrnes},\ and\ \citenamefont
  {Hartmann}}]{mohseni2023}%
  \BibitemOpen
  \bibfield  {author} {\bibinfo {author} {\bibfnamefont {Naeimeh}\ \bibnamefont
  {Mohseni}}, \bibinfo {author} {\bibfnamefont {Junheng}\ \bibnamefont {Shi}},
  \bibinfo {author} {\bibfnamefont {Tim}\ \bibnamefont {Byrnes}}, \ and\
  \bibinfo {author} {\bibfnamefont {Michael}\ \bibnamefont {Hartmann}},\
  }\bibfield  {title} {\enquote {\bibinfo {title} {Deep learning of many-body
  observables and quantum information scrambling},}\ }\href@noop {} {\bibfield
  {journal} {\bibinfo  {journal} {arXiv:2302.04621}\ } (\bibinfo {year}
  {2023})}\BibitemShut {NoStop}%
\bibitem [{\citenamefont {Torlai}\ \emph {et~al.}(2018)\citenamefont {Torlai},
  \citenamefont {Mazzola}, \citenamefont {Carrasquilla}, \citenamefont
  {Troyer}, \citenamefont {Melko},\ and\ \citenamefont {Carleo}}]{torlai2018}%
  \BibitemOpen
  \bibfield  {author} {\bibinfo {author} {\bibfnamefont {Giacomo}\ \bibnamefont
  {Torlai}}, \bibinfo {author} {\bibfnamefont {Guglielmo}\ \bibnamefont
  {Mazzola}}, \bibinfo {author} {\bibfnamefont {Juan}\ \bibnamefont
  {Carrasquilla}}, \bibinfo {author} {\bibfnamefont {Matthias}\ \bibnamefont
  {Troyer}}, \bibinfo {author} {\bibfnamefont {Roger}\ \bibnamefont {Melko}}, \
  and\ \bibinfo {author} {\bibfnamefont {Giuseppe}\ \bibnamefont {Carleo}},\
  }\bibfield  {title} {\enquote {\bibinfo {title} {Neural-network quantum state
  tomography},}\ }\href@noop {} {\bibfield  {journal} {\bibinfo  {journal}
  {Nat. Phys.}\ }\textbf {\bibinfo {volume} {14}},\ \bibinfo {pages} {447--450}
  (\bibinfo {year} {2018})}\BibitemShut {NoStop}%
\bibitem [{\citenamefont {Carrasquilla}\ \emph {et~al.}(2019)\citenamefont
  {Carrasquilla}, \citenamefont {Torlai}, \citenamefont {Melko},\ and\
  \citenamefont {Aolita}}]{carrasquilla2019}%
  \BibitemOpen
  \bibfield  {author} {\bibinfo {author} {\bibfnamefont {Juan}\ \bibnamefont
  {Carrasquilla}}, \bibinfo {author} {\bibfnamefont {Giacomo}\ \bibnamefont
  {Torlai}}, \bibinfo {author} {\bibfnamefont {Roger~G}\ \bibnamefont {Melko}},
  \ and\ \bibinfo {author} {\bibfnamefont {Leandro}\ \bibnamefont {Aolita}},\
  }\bibfield  {title} {\enquote {\bibinfo {title} {Reconstructing quantum
  states with generative models},}\ }\href@noop {} {\bibfield  {journal}
  {\bibinfo  {journal} {Nat. Mach. Intell.}\ }\textbf {\bibinfo {volume} {1}},\
  \bibinfo {pages} {155--161} (\bibinfo {year} {2019})}\BibitemShut {NoStop}%
\bibitem [{\citenamefont {Ahmed}\ \emph {et~al.}(2021)\citenamefont {Ahmed},
  \citenamefont {S\'anchez Mu\~noz}, \citenamefont {Nori},\ and\ \citenamefont
  {Kockum}}]{Ahmed2021PRL}%
  \BibitemOpen
  \bibfield  {author} {\bibinfo {author} {\bibfnamefont {Shahnawaz}\
  \bibnamefont {Ahmed}}, \bibinfo {author} {\bibfnamefont {Carlos}\
  \bibnamefont {S\'anchez Mu\~noz}}, \bibinfo {author} {\bibfnamefont {Franco}\
  \bibnamefont {Nori}}, \ and\ \bibinfo {author} {\bibfnamefont {Anton~Frisk}\
  \bibnamefont {Kockum}},\ }\bibfield  {title} {\enquote {\bibinfo {title}
  {Quantum state tomography with conditional generative adversarial
  networks},}\ }\href {\doibase 10.1103/PhysRevLett.127.140502} {\bibfield
  {journal} {\bibinfo  {journal} {Phys. Rev. Lett.}\ }\textbf {\bibinfo
  {volume} {127}},\ \bibinfo {pages} {140502} (\bibinfo {year}
  {2021})}\BibitemShut {NoStop}%
\bibitem [{\citenamefont {Zhu}\ \emph {et~al.}(2022)\citenamefont {Zhu},
  \citenamefont {Wu}, \citenamefont {Bai}, \citenamefont {Wang}, \citenamefont
  {Wang},\ and\ \citenamefont {Chiribella}}]{zhu2022}%
  \BibitemOpen
  \bibfield  {author} {\bibinfo {author} {\bibfnamefont {Yan}\ \bibnamefont
  {Zhu}}, \bibinfo {author} {\bibfnamefont {Ya-Dong}\ \bibnamefont {Wu}},
  \bibinfo {author} {\bibfnamefont {Ge}~\bibnamefont {Bai}}, \bibinfo {author}
  {\bibfnamefont {Dong-Sheng}\ \bibnamefont {Wang}}, \bibinfo {author}
  {\bibfnamefont {Yuexuan}\ \bibnamefont {Wang}}, \ and\ \bibinfo {author}
  {\bibfnamefont {Giulio}\ \bibnamefont {Chiribella}},\ }\bibfield  {title}
  {\enquote {\bibinfo {title} {Flexible learning of quantum states with
  generative query neural networks},}\ }\href@noop {} {\bibfield  {journal}
  {\bibinfo  {journal} {Nat. Commun.}\ }\textbf {\bibinfo {volume} {13}},\
  \bibinfo {pages} {6222} (\bibinfo {year} {2022})}\BibitemShut {NoStop}%
\bibitem [{\citenamefont {Karras}\ \emph {et~al.}(2019)\citenamefont {Karras},
  \citenamefont {Laine},\ and\ \citenamefont {Aila}}]{karras2019style}%
  \BibitemOpen
  \bibfield  {author} {\bibinfo {author} {\bibfnamefont {Tero}\ \bibnamefont
  {Karras}}, \bibinfo {author} {\bibfnamefont {Samuli}\ \bibnamefont {Laine}},
  \ and\ \bibinfo {author} {\bibfnamefont {Timo}\ \bibnamefont {Aila}},\
  }\bibfield  {title} {\enquote {\bibinfo {title} {A style-based generator
  architecture for generative adversarial networks},}\ }in\ \href@noop {}
  {\emph {\bibinfo {booktitle} {Proceedings of the IEEE/CVF conference on
  computer vision and pattern recognition}}}\ (\bibinfo {year} {2019})\ pp.\
  \bibinfo {pages} {4401--4410}\BibitemShut {NoStop}%
\bibitem [{\citenamefont {Karras}\ \emph {et~al.}(2020)\citenamefont {Karras},
  \citenamefont {Laine}, \citenamefont {Aittala}, \citenamefont {Hellsten},
  \citenamefont {Lehtinen},\ and\ \citenamefont {Aila}}]{karras2020analyzing}%
  \BibitemOpen
  \bibfield  {author} {\bibinfo {author} {\bibfnamefont {Tero}\ \bibnamefont
  {Karras}}, \bibinfo {author} {\bibfnamefont {Samuli}\ \bibnamefont {Laine}},
  \bibinfo {author} {\bibfnamefont {Miika}\ \bibnamefont {Aittala}}, \bibinfo
  {author} {\bibfnamefont {Janne}\ \bibnamefont {Hellsten}}, \bibinfo {author}
  {\bibfnamefont {Jaakko}\ \bibnamefont {Lehtinen}}, \ and\ \bibinfo {author}
  {\bibfnamefont {Timo}\ \bibnamefont {Aila}},\ }\bibfield  {title} {\enquote
  {\bibinfo {title} {Analyzing and improving the image quality of stylegan},}\
  }in\ \href@noop {} {\emph {\bibinfo {booktitle} {Proceedings of the IEEE/CVF
  conference on computer vision and pattern recognition}}}\ (\bibinfo {year}
  {2020})\ pp.\ \bibinfo {pages} {8110--8119}\BibitemShut {NoStop}%
\bibitem [{\citenamefont {Huang}\ \emph
  {et~al.}(2022{\natexlab{c}})\citenamefont {Huang}, \citenamefont {Kueng},
  \citenamefont {Torlai}, \citenamefont {Albert},\ and\ \citenamefont
  {Preskill}}]{huang2022provably}%
  \BibitemOpen
  \bibfield  {author} {\bibinfo {author} {\bibfnamefont {Hsin-Yuan}\
  \bibnamefont {Huang}}, \bibinfo {author} {\bibfnamefont {Richard}\
  \bibnamefont {Kueng}}, \bibinfo {author} {\bibfnamefont {Giacomo}\
  \bibnamefont {Torlai}}, \bibinfo {author} {\bibfnamefont {Victor~V}\
  \bibnamefont {Albert}}, \ and\ \bibinfo {author} {\bibfnamefont {John}\
  \bibnamefont {Preskill}},\ }\bibfield  {title} {\enquote {\bibinfo {title}
  {Provably efficient machine learning for quantum many-body problems},}\
  }\href@noop {} {\bibfield  {journal} {\bibinfo  {journal} {Science}\ }\textbf
  {\bibinfo {volume} {377}},\ \bibinfo {pages} {eabk3333} (\bibinfo {year}
  {2022}{\natexlab{c}})}\BibitemShut {NoStop}%
\bibitem [{\citenamefont {Krzywinski}\ and\ \citenamefont
  {Altman}(2014)}]{krzywinski2014visualizing}%
  \BibitemOpen
  \bibfield  {author} {\bibinfo {author} {\bibfnamefont {Martin}\ \bibnamefont
  {Krzywinski}}\ and\ \bibinfo {author} {\bibfnamefont {Naomi}\ \bibnamefont
  {Altman}},\ }\bibfield  {title} {\enquote {\bibinfo {title} {Visualizing
  samples with box plots.}}\ }\href@noop {} {\bibfield  {journal} {\bibinfo
  {journal} {Nature methods}\ }\textbf {\bibinfo {volume} {11}},\ \bibinfo
  {pages} {119--120} (\bibinfo {year} {2014})}\BibitemShut {NoStop}%
\bibitem [{\citenamefont {Lanyon}\ \emph {et~al.}(2017)\citenamefont {Lanyon},
  \citenamefont {Maier}, \citenamefont {Holz{\"a}pfel}, \citenamefont
  {Baumgratz}, \citenamefont {Hempel}, \citenamefont {Jurcevic}, \citenamefont
  {Dhand}, \citenamefont {Buyskikh}, \citenamefont {Daley}, \citenamefont
  {Cramer} \emph {et~al.}}]{lanyon2017}%
  \BibitemOpen
  \bibfield  {author} {\bibinfo {author} {\bibfnamefont {BP}~\bibnamefont
  {Lanyon}}, \bibinfo {author} {\bibfnamefont {C}~\bibnamefont {Maier}},
  \bibinfo {author} {\bibfnamefont {Milan}\ \bibnamefont {Holz{\"a}pfel}},
  \bibinfo {author} {\bibfnamefont {Tillmann}\ \bibnamefont {Baumgratz}},
  \bibinfo {author} {\bibfnamefont {C}~\bibnamefont {Hempel}}, \bibinfo
  {author} {\bibfnamefont {P}~\bibnamefont {Jurcevic}}, \bibinfo {author}
  {\bibfnamefont {Ish}\ \bibnamefont {Dhand}}, \bibinfo {author} {\bibfnamefont
  {AS}~\bibnamefont {Buyskikh}}, \bibinfo {author} {\bibfnamefont
  {AJ}~\bibnamefont {Daley}}, \bibinfo {author} {\bibfnamefont {Marcus}\
  \bibnamefont {Cramer}},  \emph {et~al.},\ }\bibfield  {title} {\enquote
  {\bibinfo {title} {Efficient tomography of a quantum many-body system},}\
  }\href@noop {} {\bibfield  {journal} {\bibinfo  {journal} {Nat. Phys.}\
  }\textbf {\bibinfo {volume} {13}},\ \bibinfo {pages} {1158--1162} (\bibinfo
  {year} {2017})}\BibitemShut {NoStop}%
\bibitem [{\citenamefont {Friis}\ \emph {et~al.}(2018)\citenamefont {Friis},
  \citenamefont {Marty}, \citenamefont {Maier}, \citenamefont {Hempel},
  \citenamefont {Holz\"apfel}, \citenamefont {Jurcevic}, \citenamefont
  {Plenio}, \citenamefont {Huber}, \citenamefont {Roos}, \citenamefont
  {Blatt},\ and\ \citenamefont {Lanyon}}]{friis2018}%
  \BibitemOpen
  \bibfield  {author} {\bibinfo {author} {\bibfnamefont {Nicolai}\ \bibnamefont
  {Friis}}, \bibinfo {author} {\bibfnamefont {Oliver}\ \bibnamefont {Marty}},
  \bibinfo {author} {\bibfnamefont {Christine}\ \bibnamefont {Maier}}, \bibinfo
  {author} {\bibfnamefont {Cornelius}\ \bibnamefont {Hempel}}, \bibinfo
  {author} {\bibfnamefont {Milan}\ \bibnamefont {Holz\"apfel}}, \bibinfo
  {author} {\bibfnamefont {Petar}\ \bibnamefont {Jurcevic}}, \bibinfo {author}
  {\bibfnamefont {Martin~B.}\ \bibnamefont {Plenio}}, \bibinfo {author}
  {\bibfnamefont {Marcus}\ \bibnamefont {Huber}}, \bibinfo {author}
  {\bibfnamefont {Christian}\ \bibnamefont {Roos}}, \bibinfo {author}
  {\bibfnamefont {Rainer}\ \bibnamefont {Blatt}}, \ and\ \bibinfo {author}
  {\bibfnamefont {Ben}\ \bibnamefont {Lanyon}},\ }\bibfield  {title} {\enquote
  {\bibinfo {title} {Observation of entangled states of a fully controlled
  20-qubit system},}\ }\href {\doibase 10.1103/PhysRevX.8.021012} {\bibfield
  {journal} {\bibinfo  {journal} {Phys. Rev. X}\ }\textbf {\bibinfo {volume}
  {8}},\ \bibinfo {pages} {021012} (\bibinfo {year} {2018})}\BibitemShut
  {NoStop}%
\bibitem [{\citenamefont {Hintze}\ and\ \citenamefont
  {Nelson}(1998)}]{hintze1998violin}%
  \BibitemOpen
  \bibfield  {author} {\bibinfo {author} {\bibfnamefont {Jerry~L}\ \bibnamefont
  {Hintze}}\ and\ \bibinfo {author} {\bibfnamefont {Ray~D}\ \bibnamefont
  {Nelson}},\ }\bibfield  {title} {\enquote {\bibinfo {title} {Violin plots: a
  box plot-density trace synergism},}\ }\href@noop {} {\bibfield  {journal}
  {\bibinfo  {journal} {The American Statistician}\ }\textbf {\bibinfo {volume}
  {52}},\ \bibinfo {pages} {181--184} (\bibinfo {year} {1998})}\BibitemShut
  {NoStop}%
\bibitem [{\citenamefont {Dykman}(2012)}]{dykman2012}%
  \BibitemOpen
  \bibfield  {author} {\bibinfo {author} {\bibfnamefont {Mark}\ \bibnamefont
  {Dykman}},\ }\href@noop {} {\emph {\bibinfo {title} {Fluctuating nonlinear
  oscillators: from nanomechanics to quantum superconducting circuits}}}\
  (\bibinfo  {publisher} {Oxford University Press},\ \bibinfo {year}
  {2012})\BibitemShut {NoStop}%
\bibitem [{\citenamefont {Temme}\ \emph {et~al.}(2017)\citenamefont {Temme},
  \citenamefont {Bravyi},\ and\ \citenamefont {Gambetta}}]{temme2017}%
  \BibitemOpen
  \bibfield  {author} {\bibinfo {author} {\bibfnamefont {Kristan}\ \bibnamefont
  {Temme}}, \bibinfo {author} {\bibfnamefont {Sergey}\ \bibnamefont {Bravyi}},
  \ and\ \bibinfo {author} {\bibfnamefont {Jay~M.}\ \bibnamefont {Gambetta}},\
  }\bibfield  {title} {\enquote {\bibinfo {title} {Error mitigation for
  short-depth quantum circuits},}\ }\href {\doibase
  10.1103/PhysRevLett.119.180509} {\bibfield  {journal} {\bibinfo  {journal}
  {Phys. Rev. Lett.}\ }\textbf {\bibinfo {volume} {119}},\ \bibinfo {pages}
  {180509} (\bibinfo {year} {2017})}\BibitemShut {NoStop}%
\bibitem [{\citenamefont {Endo}\ \emph {et~al.}(2018)\citenamefont {Endo},
  \citenamefont {Benjamin},\ and\ \citenamefont {Li}}]{endo2018}%
  \BibitemOpen
  \bibfield  {author} {\bibinfo {author} {\bibfnamefont {Suguru}\ \bibnamefont
  {Endo}}, \bibinfo {author} {\bibfnamefont {Simon~C.}\ \bibnamefont
  {Benjamin}}, \ and\ \bibinfo {author} {\bibfnamefont {Ying}\ \bibnamefont
  {Li}},\ }\bibfield  {title} {\enquote {\bibinfo {title} {Practical quantum
  error mitigation for near-future applications},}\ }\href {\doibase
  10.1103/PhysRevX.8.031027} {\bibfield  {journal} {\bibinfo  {journal} {Phys.
  Rev. X}\ }\textbf {\bibinfo {volume} {8}},\ \bibinfo {pages} {031027}
  (\bibinfo {year} {2018})}\BibitemShut {NoStop}%
\bibitem [{\citenamefont {Adesso}\ and\ \citenamefont
  {Chiribella}(2008)}]{adesso2008}%
  \BibitemOpen
  \bibfield  {author} {\bibinfo {author} {\bibfnamefont {Gerardo}\ \bibnamefont
  {Adesso}}\ and\ \bibinfo {author} {\bibfnamefont {Giulio}\ \bibnamefont
  {Chiribella}},\ }\bibfield  {title} {\enquote {\bibinfo {title} {Quantum
  benchmark for teleportation and storage of squeezed states},}\ }\href
  {\doibase 10.1103/PhysRevLett.100.170503} {\bibfield  {journal} {\bibinfo
  {journal} {Phys. Rev. Lett.}\ }\textbf {\bibinfo {volume} {100}},\ \bibinfo
  {pages} {170503} (\bibinfo {year} {2008})}\BibitemShut {NoStop}%
\bibitem [{\citenamefont {Chiribella}\ and\ \citenamefont
  {Xie}(2013)}]{chiribella2013}%
  \BibitemOpen
  \bibfield  {author} {\bibinfo {author} {\bibfnamefont {Giulio}\ \bibnamefont
  {Chiribella}}\ and\ \bibinfo {author} {\bibfnamefont {Jinyu}\ \bibnamefont
  {Xie}},\ }\bibfield  {title} {\enquote {\bibinfo {title} {Optimal design and
  quantum benchmarks for coherent state amplifiers},}\ }\href {\doibase
  10.1103/PhysRevLett.110.213602} {\bibfield  {journal} {\bibinfo  {journal}
  {Phys. Rev. Lett.}\ }\textbf {\bibinfo {volume} {110}},\ \bibinfo {pages}
  {213602} (\bibinfo {year} {2013})}\BibitemShut {NoStop}%
\bibitem [{\citenamefont {Aaronson}(2007)}]{aaronson2007}%
  \BibitemOpen
  \bibfield  {author} {\bibinfo {author} {\bibfnamefont {Scott}\ \bibnamefont
  {Aaronson}},\ }\bibfield  {title} {\enquote {\bibinfo {title} {The
  learnability of quantum states},}\ }\href@noop {} {\bibfield  {journal}
  {\bibinfo  {journal} {Proc. R. Soc. A}\ }\textbf {\bibinfo {volume} {463}},\
  \bibinfo {pages} {3089--3114} (\bibinfo {year} {2007})}\BibitemShut {NoStop}%
\bibitem [{\citenamefont {Caro}(2022)}]{caro2022}%
  \BibitemOpen
  \bibfield  {author} {\bibinfo {author} {\bibfnamefont {Matthias~C}\
  \bibnamefont {Caro}},\ }\bibfield  {title} {\enquote {\bibinfo {title}
  {Learning quantum processes and hamiltonians via the pauli transfer
  matrix},}\ }\href@noop {} {\bibfield  {journal} {\bibinfo  {journal}
  {arXiv:2212.04471}\ } (\bibinfo {year} {2022})}\BibitemShut {NoStop}%
\bibitem [{\citenamefont {Marvian}\ and\ \citenamefont
  {Lloyd}(2016)}]{marvian2016}%
  \BibitemOpen
  \bibfield  {author} {\bibinfo {author} {\bibfnamefont {Iman}\ \bibnamefont
  {Marvian}}\ and\ \bibinfo {author} {\bibfnamefont {Seth}\ \bibnamefont
  {Lloyd}},\ }\bibfield  {title} {\enquote {\bibinfo {title} {Universal quantum
  emulator},}\ }\href@noop {} {\bibfield  {journal} {\bibinfo  {journal}
  {arXiv:1606.02734}\ } (\bibinfo {year} {2016})}\BibitemShut {NoStop}%
\bibitem [{\citenamefont {Bisio}\ \emph {et~al.}(2010)\citenamefont {Bisio},
  \citenamefont {Chiribella}, \citenamefont {D'Ariano}, \citenamefont
  {Facchini},\ and\ \citenamefont {Perinotti}}]{bisio2010}%
  \BibitemOpen
  \bibfield  {author} {\bibinfo {author} {\bibfnamefont {Alessandro}\
  \bibnamefont {Bisio}}, \bibinfo {author} {\bibfnamefont {Giulio}\
  \bibnamefont {Chiribella}}, \bibinfo {author} {\bibfnamefont {Giacomo~Mauro}\
  \bibnamefont {D'Ariano}}, \bibinfo {author} {\bibfnamefont {Stefano}\
  \bibnamefont {Facchini}}, \ and\ \bibinfo {author} {\bibfnamefont {Paolo}\
  \bibnamefont {Perinotti}},\ }\bibfield  {title} {\enquote {\bibinfo {title}
  {Optimal quantum learning of a unitary transformation},}\ }\href {\doibase
  10.1103/PhysRevA.81.032324} {\bibfield  {journal} {\bibinfo  {journal} {Phys.
  Rev. A}\ }\textbf {\bibinfo {volume} {81}},\ \bibinfo {pages} {032324}
  (\bibinfo {year} {2010})}\BibitemShut {NoStop}%
\bibitem [{\citenamefont {Caro}\ \emph {et~al.}(2023)\citenamefont {Caro},
  \citenamefont {Huang}, \citenamefont {Ezzell}, \citenamefont {Gibbs},
  \citenamefont {Sornborger}, \citenamefont {Cincio}, \citenamefont {Coles},\
  and\ \citenamefont {Holmes}}]{caro2023out}%
  \BibitemOpen
  \bibfield  {author} {\bibinfo {author} {\bibfnamefont {Matthias~C}\
  \bibnamefont {Caro}}, \bibinfo {author} {\bibfnamefont {Hsin-Yuan}\
  \bibnamefont {Huang}}, \bibinfo {author} {\bibfnamefont {Nicholas}\
  \bibnamefont {Ezzell}}, \bibinfo {author} {\bibfnamefont {Joe}\ \bibnamefont
  {Gibbs}}, \bibinfo {author} {\bibfnamefont {Andrew~T}\ \bibnamefont
  {Sornborger}}, \bibinfo {author} {\bibfnamefont {Lukasz}\ \bibnamefont
  {Cincio}}, \bibinfo {author} {\bibfnamefont {Patrick~J}\ \bibnamefont
  {Coles}}, \ and\ \bibinfo {author} {\bibfnamefont {Zo{\"e}}\ \bibnamefont
  {Holmes}},\ }\bibfield  {title} {\enquote {\bibinfo {title}
  {Out-of-distribution generalization for learning quantum dynamics},}\
  }\href@noop {} {\bibfield  {journal} {\bibinfo  {journal} {Nat. Commun.}\
  }\textbf {\bibinfo {volume} {14}},\ \bibinfo {pages} {3751} (\bibinfo {year}
  {2023})}\BibitemShut {NoStop}%
\bibitem [{\citenamefont {Jerbi}\ \emph {et~al.}(2023)\citenamefont {Jerbi},
  \citenamefont {Gibbs}, \citenamefont {Rudolph}, \citenamefont {Caro},
  \citenamefont {Coles}, \citenamefont {Huang},\ and\ \citenamefont
  {Holmes}}]{jerbi2023}%
  \BibitemOpen
  \bibfield  {author} {\bibinfo {author} {\bibfnamefont {Sofiene}\ \bibnamefont
  {Jerbi}}, \bibinfo {author} {\bibfnamefont {Joe}\ \bibnamefont {Gibbs}},
  \bibinfo {author} {\bibfnamefont {Manuel~S}\ \bibnamefont {Rudolph}},
  \bibinfo {author} {\bibfnamefont {Matthias~C}\ \bibnamefont {Caro}}, \bibinfo
  {author} {\bibfnamefont {Patrick~J}\ \bibnamefont {Coles}}, \bibinfo {author}
  {\bibfnamefont {Hsin-Yuan}\ \bibnamefont {Huang}}, \ and\ \bibinfo {author}
  {\bibfnamefont {Zo{\"e}}\ \bibnamefont {Holmes}},\ }\bibfield  {title}
  {\enquote {\bibinfo {title} {The power and limitations of learning quantum
  dynamics incoherently},}\ }\href@noop {} {\bibfield  {journal} {\bibinfo
  {journal} {arXiv:2303.12834}\ } (\bibinfo {year} {2023})}\BibitemShut
  {NoStop}%
\bibitem [{\citenamefont {Genois}\ \emph {et~al.}(2021)\citenamefont {Genois},
  \citenamefont {Gross}, \citenamefont {Di~Paolo}, \citenamefont {Stevenson},
  \citenamefont {Koolstra}, \citenamefont {Hashim}, \citenamefont {Siddiqi},\
  and\ \citenamefont {Blais}}]{genois2021}%
  \BibitemOpen
  \bibfield  {author} {\bibinfo {author} {\bibfnamefont {\'Elie}\ \bibnamefont
  {Genois}}, \bibinfo {author} {\bibfnamefont {Jonathan~A.}\ \bibnamefont
  {Gross}}, \bibinfo {author} {\bibfnamefont {Agustin}\ \bibnamefont
  {Di~Paolo}}, \bibinfo {author} {\bibfnamefont {Noah~J.}\ \bibnamefont
  {Stevenson}}, \bibinfo {author} {\bibfnamefont {Gerwin}\ \bibnamefont
  {Koolstra}}, \bibinfo {author} {\bibfnamefont {Akel}\ \bibnamefont {Hashim}},
  \bibinfo {author} {\bibfnamefont {Irfan}\ \bibnamefont {Siddiqi}}, \ and\
  \bibinfo {author} {\bibfnamefont {Alexandre}\ \bibnamefont {Blais}},\
  }\bibfield  {title} {\enquote {\bibinfo {title} {Quantum-tailored
  machine-learning characterization of a superconducting qubit},}\ }\href
  {\doibase 10.1103/PRXQuantum.2.040355} {\bibfield  {journal} {\bibinfo
  {journal} {PRX Quantum}\ }\textbf {\bibinfo {volume} {2}},\ \bibinfo {pages}
  {040355} (\bibinfo {year} {2021})}\BibitemShut {NoStop}%
\bibitem [{\citenamefont {Iten}\ \emph {et~al.}(2020)\citenamefont {Iten},
  \citenamefont {Metger}, \citenamefont {Wilming}, \citenamefont {del Rio},\
  and\ \citenamefont {Renner}}]{iten2020}%
  \BibitemOpen
  \bibfield  {author} {\bibinfo {author} {\bibfnamefont {Raban}\ \bibnamefont
  {Iten}}, \bibinfo {author} {\bibfnamefont {Tony}\ \bibnamefont {Metger}},
  \bibinfo {author} {\bibfnamefont {Henrik}\ \bibnamefont {Wilming}}, \bibinfo
  {author} {\bibfnamefont {L\'{\i}dia}\ \bibnamefont {del Rio}}, \ and\
  \bibinfo {author} {\bibfnamefont {Renato}\ \bibnamefont {Renner}},\
  }\bibfield  {title} {\enquote {\bibinfo {title} {Discovering physical
  concepts with neural networks},}\ }\href {\doibase
  10.1103/PhysRevLett.124.010508} {\bibfield  {journal} {\bibinfo  {journal}
  {Phys. Rev. Lett.}\ }\textbf {\bibinfo {volume} {124}},\ \bibinfo {pages}
  {010508} (\bibinfo {year} {2020})}\BibitemShut {NoStop}%
\bibitem [{\citenamefont {Iten}(2023)}]{iten2023artificial}%
  \BibitemOpen
  \bibfield  {author} {\bibinfo {author} {\bibfnamefont {Raban}\ \bibnamefont
  {Iten}},\ }\href@noop {} {\emph {\bibinfo {title} {Artificial Intelligence
  for Scientific Discoveries: Extracting Physical Concepts from Experimental
  Data Using Deep Learning}}}\ (\bibinfo  {publisher} {Springer Nature},\
  \bibinfo {year} {2023})\BibitemShut {NoStop}%
\bibitem [{\citenamefont {Aggarwal}\ \emph {et~al.}(2018)\citenamefont
  {Aggarwal} \emph {et~al.}}]{aggarwal2018neural}%
  \BibitemOpen
  \bibfield  {author} {\bibinfo {author} {\bibfnamefont {Charu~C}\ \bibnamefont
  {Aggarwal}} \emph {et~al.},\ }\bibfield  {title} {\enquote {\bibinfo {title}
  {Neural networks and deep learning},}\ }\href@noop {} {\bibfield  {journal}
  {\bibinfo  {journal} {Springer}\ }\textbf {\bibinfo {volume} {10}},\ \bibinfo
  {pages} {3} (\bibinfo {year} {2018})}\BibitemShut {NoStop}%
\bibitem [{\citenamefont {Hochreiter}\ and\ \citenamefont
  {Schmidhuber}(1997)}]{hochreiter1997long}%
  \BibitemOpen
  \bibfield  {author} {\bibinfo {author} {\bibfnamefont {Sepp}\ \bibnamefont
  {Hochreiter}}\ and\ \bibinfo {author} {\bibfnamefont {J{\"u}rgen}\
  \bibnamefont {Schmidhuber}},\ }\bibfield  {title} {\enquote {\bibinfo {title}
  {Long short-term memory},}\ }\href@noop {} {\bibfield  {journal} {\bibinfo
  {journal} {Neural computation}\ }\textbf {\bibinfo {volume} {9}},\ \bibinfo
  {pages} {1735--1780} (\bibinfo {year} {1997})}\BibitemShut {NoStop}%
\bibitem [{\citenamefont {Kingma}\ and\ \citenamefont
  {Ba}(2014)}]{kingma2014adam}%
  \BibitemOpen
  \bibfield  {author} {\bibinfo {author} {\bibfnamefont {Diederik~P}\
  \bibnamefont {Kingma}}\ and\ \bibinfo {author} {\bibfnamefont {Jimmy}\
  \bibnamefont {Ba}},\ }\bibfield  {title} {\enquote {\bibinfo {title} {Adam: A
  method for stochastic optimization},}\ }\href@noop {} {\bibfield  {journal}
  {\bibinfo  {journal} {arXiv preprint arXiv:1412.6980}\ } (\bibinfo {year}
  {2014})}\BibitemShut {NoStop}%
\bibitem [{\citenamefont {Developers}(2022)}]{cirq_developers_2022_7465577}%
  \BibitemOpen
  \bibfield  {author} {\bibinfo {author} {\bibfnamefont {Cirq}\ \bibnamefont
  {Developers}},\ }\href {\doibase 10.5281/zenodo.7465577} {\enquote {\bibinfo
  {title} {Cirq},}\ } (\bibinfo {year} {2022}),\ \bibinfo {note} {{See full
  list of authors on Github: https://github
  .com/quantumlib/Cirq/graphs/contributors}}\BibitemShut {NoStop}%
\bibitem [{\citenamefont {Suzuki}(1976)}]{suzuki1976generalized}%
  \BibitemOpen
  \bibfield  {author} {\bibinfo {author} {\bibfnamefont {Masuo}\ \bibnamefont
  {Suzuki}},\ }\bibfield  {title} {\enquote {\bibinfo {title} {Generalized
  trotter's formula and systematic approximants of exponential operators and
  inner derivations with applications to many-body problems},}\ }\href@noop {}
  {\bibfield  {journal} {\bibinfo  {journal} {Communications in Mathematical
  Physics}\ }\textbf {\bibinfo {volume} {51}},\ \bibinfo {pages} {183--190}
  (\bibinfo {year} {1976})}\BibitemShut {NoStop}%
\bibitem [{\citenamefont {Killoran}\ \emph {et~al.}(2019)\citenamefont
  {Killoran}, \citenamefont {Izaac}, \citenamefont {Quesada}, \citenamefont
  {Bergholm}, \citenamefont {Amy},\ and\ \citenamefont
  {Weedbrook}}]{killoran2019strawberry}%
  \BibitemOpen
  \bibfield  {author} {\bibinfo {author} {\bibfnamefont {Nathan}\ \bibnamefont
  {Killoran}}, \bibinfo {author} {\bibfnamefont {Josh}\ \bibnamefont {Izaac}},
  \bibinfo {author} {\bibfnamefont {Nicol{\'a}s}\ \bibnamefont {Quesada}},
  \bibinfo {author} {\bibfnamefont {Ville}\ \bibnamefont {Bergholm}}, \bibinfo
  {author} {\bibfnamefont {Matthew}\ \bibnamefont {Amy}}, \ and\ \bibinfo
  {author} {\bibfnamefont {Christian}\ \bibnamefont {Weedbrook}},\ }\bibfield
  {title} {\enquote {\bibinfo {title} {Strawberry fields: A software platform
  for photonic quantum computing},}\ }\href@noop {} {\bibfield  {journal}
  {\bibinfo  {journal} {Quantum}\ }\textbf {\bibinfo {volume} {3}},\ \bibinfo
  {pages} {129} (\bibinfo {year} {2019})}\BibitemShut {NoStop}%
\bibitem [{\citenamefont {Paszke}\ \emph {et~al.}(2019)\citenamefont {Paszke},
  \citenamefont {Gross}, \citenamefont {Massa}, \citenamefont {Lerer},
  \citenamefont {Bradbury}, \citenamefont {Chanan}, \citenamefont {Killeen},
  \citenamefont {Lin}, \citenamefont {Gimelshein}, \citenamefont {Antiga} \emph
  {et~al.}}]{paszke2019pytorch}%
  \BibitemOpen
  \bibfield  {author} {\bibinfo {author} {\bibfnamefont {Adam}\ \bibnamefont
  {Paszke}}, \bibinfo {author} {\bibfnamefont {Sam}\ \bibnamefont {Gross}},
  \bibinfo {author} {\bibfnamefont {Francisco}\ \bibnamefont {Massa}}, \bibinfo
  {author} {\bibfnamefont {Adam}\ \bibnamefont {Lerer}}, \bibinfo {author}
  {\bibfnamefont {James}\ \bibnamefont {Bradbury}}, \bibinfo {author}
  {\bibfnamefont {Gregory}\ \bibnamefont {Chanan}}, \bibinfo {author}
  {\bibfnamefont {Trevor}\ \bibnamefont {Killeen}}, \bibinfo {author}
  {\bibfnamefont {Zeming}\ \bibnamefont {Lin}}, \bibinfo {author}
  {\bibfnamefont {Natalia}\ \bibnamefont {Gimelshein}}, \bibinfo {author}
  {\bibfnamefont {Luca}\ \bibnamefont {Antiga}},  \emph {et~al.},\ }\bibfield
  {title} {\enquote {\bibinfo {title} {Pytorch: An imperative style,
  high-performance deep learning library},}\ }\href@noop {} {\bibfield
  {journal} {\bibinfo  {journal} {Advances in neural information processing
  systems}\ }\textbf {\bibinfo {volume} {32}} (\bibinfo {year}
  {2019})}\BibitemShut {NoStop}%
\end{thebibliography}%

\begin{appendix}
\section{Relation with previous works}

Recently, the approach of quantum shadow tomography~\cite{aaronson2007,aaronson2018,huang2020} was extended to  predicting the expectation values of observables at the  output states of an unknown quantum process~\cite{huang2022}.   A strength of this method is that it can be applied to arbitrary quantum processes. The flipside of this generality, however, is that the experimenter needs to perform uniformly randomized measurements from an informationally complete set (e.g. the set of Pauli or Clifford measurements) and to test the process on an informationally complete set of states. 
Different from Ref.~\cite{huang2022}, our approach does not require the ensemble of accessible input states or the set of accessible measurements  to be informationally complete. This flexibility is practically relevant in scenarios where the experimenter can only prepare a subset of the possible quantum states and perform a subset of the possible measurements---a situation that frequently arises in intermediate-scale quantum systems.
For informationally incomplete sets, our neural network produces a lower-dimensional description of the process, which is different from the full quantum description or from the Pauli transfer matrix used in Ref.~\cite{caro2022}. This approach works  when the input and output quantum states have sufficiently regular structure, which enables a compressed, data-driven state representation. 

Although there have been works on utilizing machine learning tools for predicting future observations in a quantum dynamical process using past measurement data~\cite{flurin2020,mohseni2022,koolstra2022,huangYulei2022,mohseni2023}, they only learn a quantum dynamical process with respect to a fixed initial state. Instead, the neural network we develop learns the action of a quantum process applied on various input states chosen out of a certain ensemble and makes predictions on all outputs corresponding to those inputs. 
It is also worth mentioning that our approach does not require full control over the input states,  which is instead necessary in  other quantum process learning approaches.
In fact, the experimenter may not even have access to the complete description of the input states, but only to  measurement data obtained from experiments.  These data are nevertheless sufficient to generate a representation of the unknown process, and, if the states are sufficiently regular, to produce good predictions.  

In our neural network,  a lower-dimensional representation of the process is built by comparing the measurement statistics from pairs of  input and output quantum states. This scenario is somewhat reminiscent of the task of quantum emulation \cite{marvian2016}, where a quantum circuit is designed to learn how to imitate an unknown quantum process by observing its inputs and the corresponding outputs.  Unlike in quantum emulation, however, our network does not require direct access to the input and output states, but only to the measurement data generated from them. Moreover, the output of our network is a prediction on measurement statistics, rather than being a quantum circuit that emulates the unknown process.  This fact  is also a difference with other approaches to quantum process learning, such as ~\cite{bisio2010,caro2023out,jerbi2023}.

It is worth noting that our neural network does not 
 require any prior information on the quantum process or on the ensemble of input states, or even any  knowledge about quantum  physics.  This fact differentiates our model from other approaches that include an incorporated physical knowledge in the design of neural network models~\cite{genois2021}.   

\section{Neural Network Structure}
\label{sec:neuralNetwork}



In this section, we are going to explain the structure of the neural network model.
 The deep neural network model for learning a quantum process is a combination of a representation network, a neural emulator and a generation network.
The combination of representation network that produces a data-driven state representation using the measurement data and generation network that uses a state representation to produce the predictions of the measurement statistics for any query measurement, is called a generative query neural network for quantum (GQNQ), which has been introduced in Ref.~\cite{zhu2022} for the purpose of learning quantum states.
The neural emulator maps an input state representation to an output state representation, which is a low-dimensional analogue of the quantum process $\mathcal E$, where the dimension $d$ of the state representations can be much smaller than $\text{dim}(\mathcal H)$. A higher-dimensional state representation empirically yields better predictive performance but also leads to trickier training process.

Now we first explain how our neural model works in the prediction phase and then discuss how we establish such a neural model in the training phase.
In the prediction phase, given any input state $\rho\in\mathcal S$, the experimenter calculates the measurement statistics $\bm{p}_j$ of $\rho$ for every measurement $\bm{M}_j\in \mathcal M$. The experimenter provides the learner with the data $\{(\bm{m}_j, \bm{p}_j)\}_{j=1}^{|\mathcal M|}$. 
For each pair of $(\bm{m}_j, \bm{p}_j)$, the representation network $f_{\bm{\xi}}$, with parameter $\bm{\xi}$, produces a $d$-dimensional representation $\bm{r}_j:=f_{\bm{\xi}}(\bm{m}_j,\bm{p}_j)$.
 Then with the set $\{\bm{r}_j\}_{j=1}^{|\mathcal M|}$ as input, an aggregation function produces the input state representation $\bm{r}_\rho$. Here we use the average as the aggregation function, i.e., $\bm{r}_\rho:=1/|\mathcal M|\sum_{j=1}^{|\mathcal M|} \bm{r}_j \in \mathbb R^d$ is the data-driven state representation of $\rho$. 

Then a neural emulator $h_{\bm{\lambda}}^{\mathcal E}$ of quantum process $\mathcal E$ with parameters $\bm{\lambda}$ is applied on $\bm{r}_{\rho}$ to yield $h_{\bm{\lambda}}^{\mathcal E}(\bm{r}_\rho)\in \mathbb R^d$, which is regarded as the state representation of $\mathcal E(\rho)$. 
After that, for any measurement $\bm{M}'\in\mathcal M$, generation network $g_{\bm{\eta}}$ with parameter $\bm{\eta}$ takes $h_{\bm{\lambda}}^{\mathcal E}(\bm{r}_\rho)$ and measurement parametrization $\bm{m}'$ as input,
producing $g_{\bm{\eta}}\left(h_{\bm{\lambda}}^{\mathcal E}(\bm{r}_\rho),\bm{m}'\right) $ as a prediction of the true outcome statistics $\bm{p}':=\tr(\mathcal E(\rho) \bm{M}')$.

Next, let us explain how the learner optimizes the parameters $\bm{\xi}$, $\bm{\eta}$ and $\bm{\lambda}$ of the neural network model using the dataset $\mathcal D$ in the training phase. Note that our neural model is a combination of three individual networks, which can be trained either separately or jointly and combined together afterwards. Here, $\bm{\xi}$ and $\bm{\eta}$ are first jointly optimized, after which, $\bm{\lambda}$ is optimized separately. First, the learner trains the combination of representation network $f_{\bm{\xi}}$ and generation network $g_{\bm{\eta}}$.
The training of these two neural networks is carried out using the data of measurement parametrization and statistics over a fiducial set of states $\left\{\sigma_i, \mathcal E(\sigma_i)\right\}_{i=1}^n$, composed of all the pairs of input and output states in the learning phase. 
The parameters $\bm{\xi}$ and $\bm{\eta}$ are jointly optimized to minimize the average difference between predicted outcome distributions and the distributions obtained from calculation or experiments, averaged over all the fiducial states and all the query measurements. Implementation details of the representation network and the generation network can be found in Appendix.~\ref{app:RGN}.


Second, the neural emulator $h_{\bm{\lambda}}^{\mathcal E}$ is going to be trained to approximate the function mapping each input state representation to its corresponding output state representation.
We first use the trained representation network $f_{\bm{\xi}}$, together with training dataset $\mathcal D$, to produce the input state representations $\bm{r}_{\sigma_i}=1/|\mathcal M|\sum_{j=1}^{|\mathcal M|} f_{\bm{\xi}}\left(\bm{m}_j, \tr(\sigma_i\bm{M}_j)\right)$, as well as output state representations $\bm{r}_{\mathcal E(\sigma_i)}=1/|\mathcal M|\sum_{j=1}^{|\mathcal M|} f_{\bm{\xi}}\left(\bm{m}_j, \bm{p}_{ij}\right)$.
Then $\bm{\lambda}$ is optimized to minimize the loss function 
\begin{equation}\label{eqn:lossfunc}
    \sum_{i=1}^n ||h_{\bm{\lambda}}^{\mathcal E}(\bm{r}_{\sigma_i})-\bm{r}_{\mathcal E(\sigma_i)}||_{L_2}^2, 
\end{equation}
where $||\cdot||_{L_2}$ is the $L_2$ norm. 
After training is concluded, the trained network $h_{\bm{\lambda}}^{\mathcal E}$ is regarded as our classical model of quantum process $\mathcal E$. Implementation details of the neural emulator are provided in Appendix.~\ref{app:NE}.


We can compare our quantum process learning model with conventional quantum process tomography. 
In conventional quantum process tomography, the classical model reconstructed by fitting experimental data in the learning phase is a classical description of the quantum process, such as its Choi state, Kraus operators, or Pauli transfer matrix, using which, the density matrix of any input state can be mapped to the density matrix of the corresponding output state.
Since GQNQ is not constrained to any specific choice of state representation, this freedom enables the network to construct lower-dimensional representations of an ensemble of input states and their corresponding output states with sufficiently regular structure.
Here we use a neural emulator to replace the role of the classical description of the quantum process for predicting its behavior on different inputs.
Another essential difference is that we do not provide the neural network model with any hard-coded physical knowledge, similar to discovery of physical knowledge with a neural network~\cite{iten2020,iten2023artificial}.

\section{Implementation details of representation network and generation network}
\label{app:RGN}

\subsection{Structure} 
\begin{figure}[hbtp]
    \includegraphics[width=0.4\textwidth]{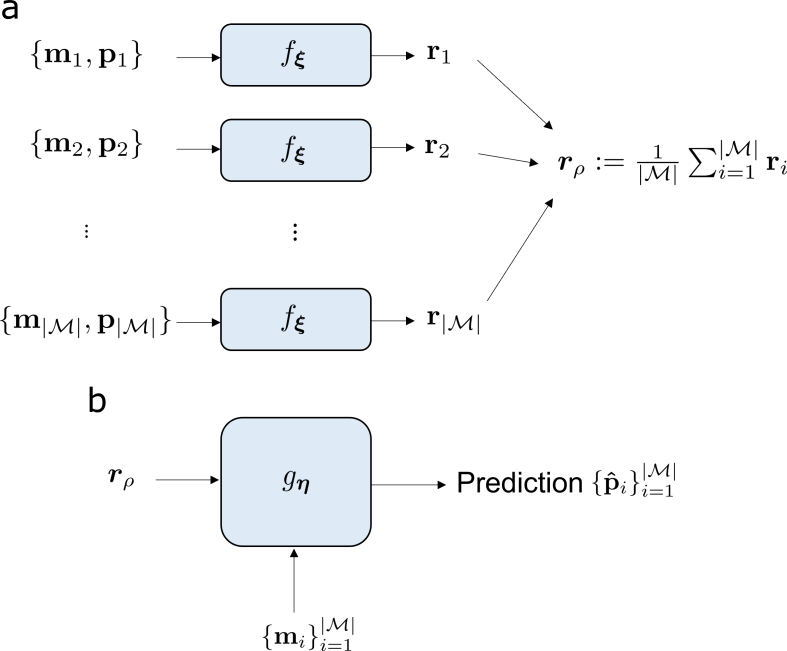}
    \caption{a. The representation network $f_{\bm{\xi}}$. b. The generation network $g_{\bm{\eta}}$.}
    \label{fig:GQNQ}
\end{figure}

We depict the structure of the representation network $f_{\bm{\xi}}$ and the generation network $g_{\bm{\eta}}$ in Fig.~\ref{fig:GQNQ}. We adopted the identical internal structure for both the representation network and generation network as presented in Ref.~\cite{zhu2022}. The representation network consists of multiple dense layers~\cite{aggarwal2018neural} and the generation network is mainly composed of  multiple dense layers and long short-term memory (LSTM) cells~\cite{hochreiter1997long}.  

We feed each pair of $(\bm{m}_i, \bm{p}_i)$ from the set $\{(\bm{m}_i, \bm{p}_i)\}_{i=1}^{|\mathcal M|}$ into the representation network to obtain the state representation of the quantum state, denoted as $r_\rho$. For the generation network, we provide both the state representation $r_\rho$ and the query measurements $\{\bm{m}_i\}_{i=1}^{|\mathcal M|}$ as inputs and expect it to generate the corresponding predictions of measurement probability distributions, denoted as $\{\bm{p}'_i\}_{i=1}^{|\mathcal M|}$.

\subsection{Training}
In the first stage of learning phase, we train the representation network and the generation network jointly with the measurement statistics over a fiducial set of states $\left\{\sigma_i, \mathcal E(\sigma_i)\right\}_{i=1}^n$ which has $2n$ states. We denote these statistics as $\{\{(\bm{m}_j, \bm{p}_j^k)\}_{j=1}^{|\mathcal M|}\}_{k=1}^{2n}$. We utilize the loss function $\mathcal{L}(\bm{\xi},\bm{\eta})$ described in Eq.~(1) of Ref.~\cite{zhu2022}. We utilize the Adam optimizer to perform batch stochastic gradient descent for minimizing the loss function $\mathcal{L}(\bm{\xi},\bm{\eta})$ during the training of both the representation network and the generation network. In all of our experiments, we maintain a fixed batch size of 5. The learning rate, initially set to 0.01, gradually decreases as the number of training epochs increases. Below is the pseudocode for Algorithm \ref{algo:training_GQNQ}, presenting the training procedure of the representation network and the generation network during the first stage of the learning phase. The notations used in the algorithm are as introduced before.

\begin{algorithm}[hbtp]
\caption{Training of representation network and generation network.}\label{algo:training_GQNQ}
\KwData{state measurement statistics $\{\{(\bm{m}_j, \bm{p}_j^k)\}_{j=1}^{|\mathcal M|}\}_{k=1}^{2n}$ corresponding to $\left\{\sigma_i, \mathcal E(\sigma_i)\right\}_{i=1}^n$, maximum number of epochs $E$, learning rate $\delta$, batch size $B$.}

Initialize parameters $\bm{\xi}$ and $\bm{\eta}$ randomly, $e = 0$\;
\While{$e<E$}{
    $\mathcal{L} = 0$\;
    \For{$k=1$ \KwTo $B$}{
         Generate a random integer number $k'$ from $[1,2n]$\;
        Input each of $\{(\bm{m}_{j}, \bm{p}^{k'}_{j})\}_{j=1}^{|\mathcal M|}$ into the representation network $f_{\bm{\xi}}$  to obtain the representations  $\{\bm{r}_{j}\}_{j=1}^{|\mathcal M|}$ as $\bm{r}_{j} =f_{\bm{\xi}}(\bm{m}_{j}, \bm{p}^{k'}_{j})$  \;
        Calculate the state representation by $\bm{r} = \sum_{j=1}^{|\mathcal M|}\bm{r}_{j}$ \;
        Input $\bm{r}$ and $\{\bm{m}_{j}\}_{j=1}^{|\mathcal M|}$ into the generation network $g_{\bm{\eta}}$ to obtain the predictions $\{\bm{\hat{p}}^{k'}_{j}\}_{j=1}^{|\mathcal M|}$ of measurement outcome distributions  as $\bm{p}^{k'}_{j} = g_{\bm{\eta}}(\bm{r}, \bm{m}_{j})$\;
        Calculate the loss $l$ by comparing $\{\bm{\hat{p}}^{k'}_{j}\}_{j=1}^{|\mathcal M|}$ with $\{\bm{p}^{k'}_{j}\}_{j=1}^{|\mathcal M|}$ and update $\mathcal{L}$ as $\mathcal{L} = \mathcal{L} + l$ \;
        }
        
    Calculate $\nabla_{\bm{\xi}} \mathcal{L}$ and $\nabla_{\bm{\eta}} \mathcal{L}$ \;
    Update $\bm{\xi}$ and $\bm{\eta}$ as $\bm{\xi} = {\rm Adam}(\bm{\xi}, \nabla_{\bm{\xi}} \mathcal{L},\delta)$, $\bm{\eta} = {\rm Adam}(\bm{\eta}, \nabla_{\bm{\eta}} \mathcal{L},\delta)$  \;
    $\mathcal{L} = 0$\;
    $e = e + 1$ \; 
    } 
\end{algorithm}

\section{Implementation details of neural emulator}
\label{app:NE}
\subsection{Structure}
\begin{figure}[hbtp]
    \includegraphics[width=0.15\textwidth]{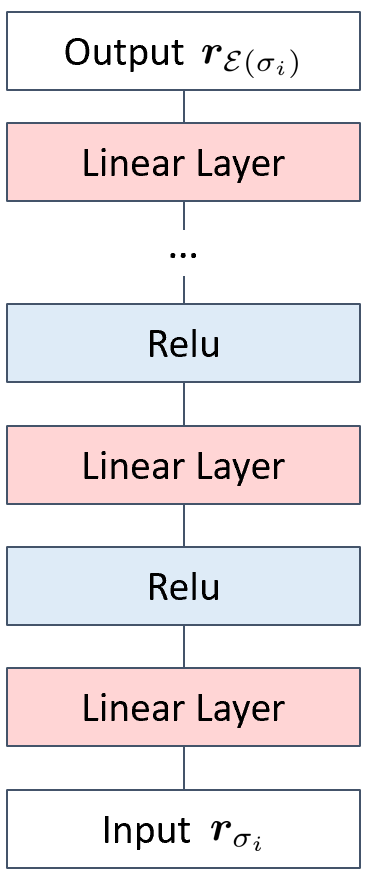}
    \caption{The neural emulator $h_{\bm{\lambda}}$.}
    \label{fig:NE}
\end{figure}

We depict the architectural design of the neural emulator in  Fig.~\ref{fig:NE}. The input to the neural emulator is the state representation of the input state of the process being learned while the output of the neural emulator is the predicted state representation of the output state. In other words, the neural emulator takes in the encoded information of the input state and generates a corresponding classical representation of the expected output state. As for the detailed architecture, the neural emulator comprises several linear layers and utilizes the ReLU activation function~\cite{aggarwal2018neural}. It is important to note that the final linear layer does not have a ReLU activation function following it, as this design choice has been shown to improve performance.

\subsection{Training}
In the second stage of the learning phase, we proceed to train the neural emulator using the state representations of pairs consisting of an input state and its corresponding output state from the fiducial set. We utilize the loss function described in Equation~\ref{eqn:lossfunc}  and employ the Adam optimizer~\cite{kingma2014adam} to perform batch gradient descent in order to minimize this loss function. We present the training procedure by for neural emulator pseudocode in Algorithm~\ref{algo:training_NE}.

\begin{algorithm}[hbtp]
\caption{Training of neural emulator.}\label{algo:training_NE}
\KwData{state representations $\{\{(\bm{r}_{\sigma_i}, \bm{r}_{\mathcal E(\sigma_i)_i})\}\}_{k=1}^{n}$ corresponding to $\left\{\sigma_i, \mathcal E(\sigma_i)\right\}_{i=1}^n$, maximum number of epochs $E$, learning rate $\delta$, batch size $B$.}

Initialize parameters $\bm{\lambda}$ and $\bm{\eta}$ randomly, $e = 0$\;
\While{$e<E$}{
    $\mathcal{L} = 0$\;
    \For{$i=1$ \KwTo $n$}{
        Input $\bm{r}_{\sigma_i}$ into the neural emulator $h_{\bm{\lambda}}^{\mathcal E}$ to obtain the predicted state representation of the output state $E(\sigma_i)_i$  as $h_{\bm{\lambda}}^{\mathcal E}(\bm{r}_{\sigma_i})$\;
        Calculate the loss $l$ by Equation~\ref{eqn:lossfunc} and update $\mathcal{L}$ as $\mathcal{L} = \mathcal{L} + l$ \;
        \If{$k\ \text{mod}\ B = 0$}{
            Calculate $\nabla_{\bm{\lambda}} \mathcal{L}$ \;
            Update $\bm{\lambda}$ as $\bm{\lambda} = {\rm Adam}(\bm{\lambda}, \nabla_{\bm{\lambda}} \mathcal{L},\delta)$\;
            $\mathcal{L} = 0$\;
        }
        }
    $e = e + 1$ \; 
    } 
\end{algorithm}

\section{Data generation}
\label{app:data}
\paragraph{Multi-qubit quantum circuits.} 
We use cirq~\cite{cirq_developers_2022_7465577} to simulate the multi-qubit variational circuits. Each random single-qubit rotations is achieved using a combination of three gates: $R_x(\theta_1)$, $R_y(\theta_2)$, and $R_y(\theta_3)$. The rotation angles $\theta_1$, $\theta_2$, and $\theta_3$ are randomly chosen from the interval $[0, 2\pi]$.
\paragraph{Spin-system dynamics.} We employ an exact method to solve the Hamiltonian evolution of a six-qubit long-range XY model with a transverse field. This exact method allows us to accurately compute the time evolution of the quantum system governed by the given Hamiltonian.

\begin{figure}[hbtp]
    \includegraphics[width=0.45\textwidth]{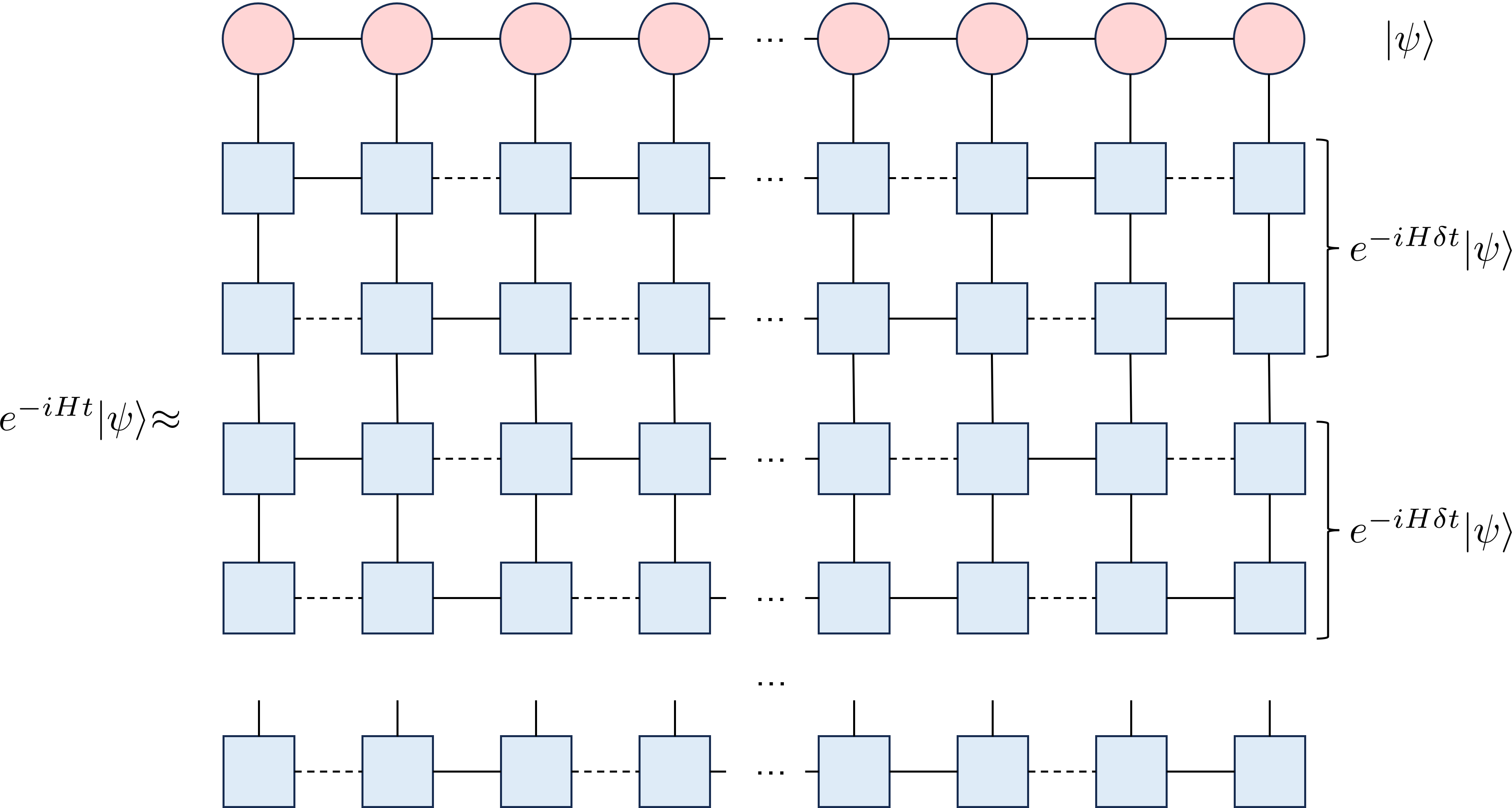}
    \caption{TEBD algorithm.}
    \label{fig:TEBD}
\end{figure}

For large-scale systems, we utilize the time-evolving block decimation (TEBD) algorithm, which is well-suited for handling systems with a large number of components. By utilizing the Trotter-Suzuki approximation~\cite{suzuki1976generalized}, the TEBD algorithm is able to efficiently capture the time evolution of the system while keeping computational complexity at a reasonable level. This enables us to study and analyze the dynamics of the large-scale system accurately. We present the TEBD algorithm we employ for the simulation in Fig.~\ref{fig:TEBD}. In our numerical experiments, we always set $\delta t$ as $0.02$.
\paragraph{Continuous-variable quantum processes.} The data of continuous-variable quantum processes are generated using simulation tools provided by Strawberry Fields~\cite{killoran2019strawberry}.

\section{Details of experiments}

\paragraph{Framework and hardware.} Our neural model is implemented using the PyTorch~\cite{paszke2019pytorch} framework and trained using the computational power of four NVIDIA GeForce GTX 1080 Ti GPUs.
\paragraph{Initialization and learning rate.} For each task, we initialize the parameters of the neural model with random values before commencing the training process. The learning rate, typically set to $0.01$ initially, is gradually reduced as the number of iterations or epochs increases. 
\paragraph{Number of epochs and training time.} Typically, we set the maximum number of epochs, denoted as $E$, to $100$ for our experiments. The actual training time may vary depending on the size of the training set for each specific task. However, the training time remains consistently under three hours for all of our experiments. 

\end{appendix}

\end{document}